\documentclass[aps,nofootinbib,preprintnumbers]{revtex4}
\usepackage{tikz}
\usepackage{lipsum}
\usepackage{CJK}
\usepackage{amsfonts}
\usepackage{amsmath}
\usepackage{amssymb}
\usepackage[english]{babel}
\usepackage{graphicx}
\usepackage{epsfig}
\usepackage{bm}
\usepackage{verbatim}
\usepackage[utf8]{inputenc}
\usepackage{booktabs}
\usepackage{multirow}
\usepackage{subfig}
\usepackage{slashed}
\usepackage{xcolor}
\usepackage[colorlinks=true,urlcolor=red,citecolor=red]{hyperref}
\usepackage[font=small]{caption}
\usepackage{float}
\usepackage{blindtext}
\usepackage{placeins}
\usepackage[utf8]{inputenc}
\usepackage{bbm}
\usepackage[colorinlistoftodos]
{todonotes}

\newenvironment{Eqnarray}{\arraycolsep 0.14em\begin{eqnarray}}{\end{eqnarray}}
\newcommand{\ba}{\begin{Eqnarray}}
\newcommand{\ea}{\end{Eqnarray}}
\newcommand{\be}{\begin{equation}}
\newcommand{\ee}{\end{equation}}
\newcommand{\bal}{\begin{aligned}}
\newcommand{\eal}{\end{aligned}}
\newcommand{\bea}{\begin{eqnarray}}
\newcommand{\eea}{\end{eqnarray}}
\newcommand{\ben}{\begin{enumerate}}
\newcommand{\een}{\end{enumerate}}
\newcommand{\bit}{\begin{itemize}}
\newcommand{\eit}{\end{itemize}}
\newcommand{\bde}{\begin{widetext}}
\newcommand{\ede}{\end{widetext}}

\def\lsim{\mathrel{\rlap{\lower4pt\hbox{\hskip1pt$\sim$}}
    \raise1pt\hbox{$<$}}}
\def\gsim{\mathrel{\rlap{\lower4pt\hbox{\hskip1pt$\sim$}}
    \raise1pt\hbox{$>$}}}

\def\3211{$\mathrm{SU(3) \otimes SU(2)_L \otimes U(1)_R \otimes U(1)_{B-L}}$ }
\def\321{$\mathrm{SU(3) \otimes SU(2) \otimes U(1)}$ }
\def\422{$\mathrm{SU(4) \otimes SU(2) \otimes SU(2)_R}$ }

\newcommand{\mathsym}[1]{{}}

\newcommand{\VKN}[1]{{\color{black}#1}}

\DeclareUnicodeCharacter{2212}{-}

\newcommand\blfootnote[1]{%
  \begingroup
  \renewcommand\thefootnote{}\footnote{#1}%
  \addtocounter{footnote}{-1}%
  \endgroup
}

\input{tcilatex}

\begin{document}
\title{A common framework for fermion mass hierarchy, leptogenesis and dark matter}

\author{Carolina Arbeláez$^{1,3~\ast}$, A.E. Cárcamo Hernández$^{1,2,3~\dagger}$, Claudio Dib$^{1,3~\ddagger}$,\\ Patricio Escalona Contreras$^{1,2~\bullet}$, Vishnudath K. N.$^{1~\diamond}$, and Alfonso Zerwekh$^{1,2,3~\odot}$}

\affiliation{$^{1}$Depto. de Fisica, Universidad T\'ecnica Federico Santa Mar\'ia, Avenida España 1680, Valpara\'iso, Chile\\ $^{2}$Millennium Institute for Subatomic Physics at High Energy Frontier – SAPHIR, Fernandez Concha 700, Santiago, Chile\\  $^{3}$Centro Cient\'ifico-Tecnol\'ogico de Valpara\'iso, Universidad T\'ecnica Federico Santa Mar\'ia, Avenida España 1680, Valpara\'iso, Chile}

\blfootnote{$^{\ast}$carolina.arbelaez@usm.cl, $^{\dagger}$antonio.carcamo@usm.cl, $^{\ddagger}$claudio.dib@usm.cl, $^{\bullet}$ patricioescalona96@gmail.com, \\ $^{\diamond}$ vishnudath.neelakand@usm.cl, $^{\odot}$ alfonso.zerwekh@usm.cl}

%\author{Carolina Arbel\'aez}
%\email{carolina.arbelaez@usm.cl}
%\affiliation{\AddrUSM}
%\affiliation{\CCTVAL}
%\author{A. E. C\'arcamo Hern\'andez}
%\email{antonio.carcamo@usm.cl}
%\affiliation{\AddrUSM}
%\affiliation{\CCTVAL}
%\affiliation{\SAPHIR}
%\author{Claudio Dib}
%\email{claudio.dib@usm.cl}
%\affiliation{\AddrUSM}
%\affiliation{\CCTVAL}
%\author{Patricio Escalona Contreras}
%\email{patricioescalona96@gmail.com}
%\affiliation{\AddrUSM}
%\affiliation{\SAPHIR}
%\author{Vishnudath K. N.}
%\email{vishnudath.neelakand@usm.cl}
%\affiliation{\AddrUSM}
%\author{Alfonso Zerwekh}
%\email{alfonso.zerwekh@usm.cl}
%\affiliation{\AddrUSM}
%\affiliation{\CCTVAL}
%\affiliation{\SAPHIR}

\begin{abstract}
 In this work, we explore an extension of the Standard Model designed to elucidate the fermion mass hierarchy, account for the dark matter relic abundance, and explain the observed matter-antimatter asymmetry in the universe. Beyond the Standard Model particle content, our model introduces additional scalars and fermions. Notably, the light active neutrinos and the first two generations of charged fermions acquire masses at the one-loop level. The model accommodates successful low-scale leptogenesis, permitting the mass of the decaying heavy right-handed neutrino to be as low as 10 TeV. We conduct a detailed analysis of the dark matter phenomenology and explore various interesting phenomenological implications. These include charged lepton flavor violation, muon and electron anomalous magnetic moments, constraints arising from electroweak precision observables, and implications for collider experiments.
\end{abstract}

\maketitle

%\begin{flushright}
%Dedicated to the memory of Eduardo Pont\'on
%\end{flushright}

%%%%%%%%

%%%%%%%%

\section{Introduction}

%{\bf{Points to highlight}

%\begin{itemize}
%\item Minimal extension of scotogenic/2HDM - wide range of phenomenological prospects
%\item Fermion mass hierarchy
%\item having 3 RHN in the scotogenic set up $\rightarrow$ low scale leptogenesis \cite{}
%\item Rich DM phenomenology
%\end{itemize}}

 Despite the remarkable success of the Standard Model (SM) in
 explaining a wide range of experimental observations, several issues still remain unexplained. One of the main issues to which the SM does not have an answer is the non-zero light active neutrino masses as indicated by various oscillation experiments~\cite{McDonald:2016ixn,Kajita:2016cak}.
{Moreover}, the huge hierarchy in the  fermion mass spectrum, which spreads over 13 orders of magnitude from the light active neutrino mass scale up to the top quark mass, is not explained by the SM. Besides that, the pattern of quark and lepton mixings are substantially different. In the quark sector, all three mixing angles are small whereas in the lepton sector two of the mixing angles are large and one is small. The SM also does not explain the current amount of matter-antimatter asymmetry observed in the Universe~\cite{Planck:2018vyg} as well as the measured dark matter (DM) relic abundance~\cite{Bertone:2004pz}. 
All these issues suggest to consider the SM as the low energy limit of an unknown underlying
 theory. Such an underlying
 theory should be capable of successfully accommodating the observed DM relic density as well as the matter-antimatter asymmetry of the Universe and should include a dynamical mechanism responsible for the observed pattern of the SM fermion masses and mixings.

  Here in this work we propose an extension of the Inert Doublet Model (IDM)~\cite{Deshpande:1977rw}, where the scalar content is enlarged {from the IDM} by the inclusion of two electrically neutral scalar singlets and the fermion sector is augmented by adding right handed Majorana neutrinos and {charged fermions that are vector-like with respect to the SM gauge group}. The SM gauge symmetry is supplemented by the inclusion of {a} spontaneously broken discrete $Z_4$ symmetry. 
The charge assignments of the particles forbid tree level masses for the light neutrinos as well as the first two generations of SM charged fermions. However, the $Z_4$ symmetry is broken down to {a} preserved $Z_2$ symmetry  and active neutrinos  as well as the first and second generations of SM charged fermions obtain their masses radiatively at one-loop level. The light active neutrinos are Majorana particles in our {setup}. The third generation of SM charged fermions obtain tree level masses as in the SM. {Also}, out of the four scalar representations, two remain odd under the remnant $Z_2$ symmetry and thus providing a stable DM candidate. These dark scalars also contribute to the one-loop generation of the masses of the fermions, thus connecting DM with neutrino mass generation and fermion mass hierarchy. Thus, the model also generalises the features of the scotogenic model~\cite{Tao:1996vb}.
%{Moreover}, the {decays} of the heavy Majorana right handed neutrinos can explain the observed baryon asymmetry of the universe via leptogenesis~\cite{Fukugita:1986hr}.
{Moreover, the model can also accommodate successful  leptogenesis - a scenario in which the CP violating decays of the heavy Majorana right handed neutrinos into the doublet scalar and the SM leptons produce a lepton asymmetry in the early universe~\cite{Fukugita:1986hr}.
The lepton asymmetry thus generated then gets converted into baryon asymmetry via non-perturbative sphaleron processes in the early universe~\cite{Kolb:1990vq}. We find that  successful leptogenesis is viable in the model for heavy Majorana right handed neutrinos as light as 10 TeV.} Note that several extensions of the IDM theory have been proposed to explain the tiny masses of the active neutrinos via the implementation of {a} radiative seesaw mechanism at one loop level \cite{Balakrishna:1988ks, Ma:1988fp, Ma:1989ys, Ma:1990ce,
Ma:1998dn, Ma:2006km, Gu:2007ug, Ma:2008cu, Hirsch:2013ola, Aranda:2015xoa, Restrepo:2015ura, Longas:2015sxk, Fraser:2015zed, Fraser:2015mhb, Wang:2015saa, Arbelaez:2016mhg, vonderPahlen:2016cbw, Nomura:2016emz, Kownacki:2016hpm, Nomura:2017emk, Nomura:2017vzp, Bernal:2017xat, Wang:2017mcy, Bonilla:2018ynb, Calle:2018ovc, Avila:2019hhv, Alvarado:2021fbw, Arbelaez:2022ejo, Cepedello:2022xgb, CarcamoHernandez:2022vjk, Leite:2023gzl}. {In this work, we extend the loop generation of masses to the charged fermionic sector as well, thereby providing a plausible explanation for the fermion mass hierarchy.}

 In addition to the above mentioned features, our model can also explain the observed muon and electron anomalous magnetic moments. It has been known that the experimentally measured
anomalous magnetic moments ($g-2$) of both the muon and electron differ by a few standard deviations from the SM
predictions. The longstanding non-compliance of the muon $g-2$ with the SM was
first observed by {the experiment E821 at BNL}~\cite{Bennett:2006fi}
and has been recently confirmed by the Muon $g-2$ experiment at FERMILAB 
\cite{Abi:2021gix}. The discrepancy of the electron {$g-2$} with the SM prediction was revealed more recently, following an accurate measurement of the fine
structure constant~\cite{Parker:2018vye}. The different magnitudes
of the electron and the muon {$g-2$} deviations do not find an explanation within
the context of the SM and {have} motivated 
theories with extended symmetries and particle {spectra;} see ~\cite%
{Athron:2021iuf} for a very recent review.

The rest of the paper is organised as follows. In Section-\ref{model}, we provide a comprehensive overview of the model, detailing its scalar and fermionic spectrum, along with an in-depth exploration of how it generates the observed SM fermion mass hierarchy. Section-\ref{pheno} is dedicated to an extensive examination of the model's phenomenological implications, encompassing topics such as charged lepton flavor violation, anomalous magnetic moments of electron and muon, constraints derived from electroweak precision observables, collider bounds, leptogenesis, and the impact of current dark matter constraints on our model's parameter space. Our findings and overall conclusions are summarized in Section-\ref{conclusions}.

\section{The model}
\label{model}

We formulate an extension of the IDM where the
active neutrino masses as well as the first and second generation of SM charged fermion masses are radiatively generated at one loop level, whereas the masses of the third generation of SM charged
fermions arise at tree level. {The SM charged fermions of the third families} obtain their masses as in the SM from the following {Yukawa interactions}:
\begin{equation}
\overline{q}_{3L}\widetilde{\phi }u_{iR},\hspace{1cm}\overline{q}_{3L}\phi
d_{iR},\hspace{1cm}\overline{l}_{iL}\phi e_{3R},\hspace{1cm}i=1,2,3,
\label{op1}
\end{equation}
where the $SU(2)_{L}$ fermionic doublets are defined as
\begin{equation}
q{_{iL}=\left( u_{iL},d_{iL}\right) },\hspace{1cm}l_{iL}={\left( \nu
_{iL},e_{iL}\right) },\hspace{1cm}i=1,2,3,  \label{op2}
\end{equation}
and the scalar doublet $\phi$ corresponds to the SM Higgs field {and can be expanded as
\begin{equation}
\phi =\left( 
\begin{array}{c}
\phi ^{+} \\ 
\frac{1}{\sqrt{2}}\left( v+\phi _{R}^{0}+i\phi _{I}^{0}\right) 
\end{array}%
\right) .  \label{SMscalardoublet}
\end{equation}
Here, $v=246$ is the scale of spontaneous breaking of the $SU(2)_L\times U(1)_Y$ gauge symmetry.} The Yukawa couplings of the first two generations of charged fermions to the SM Higgs are forbidden by a discrete $Z_4$ symmetry. \VKN{The $Z_4$ transformation of a given field $f$ is defined as $f\rightarrow (+i)^{Q_{Z_4}} f$, where $Q_{Z_4}$ is the $Z_4$ charge listed in Tables \ref{fermions} and \ref{scalars}.} To generate one-loop level masses for the first and second {generations} of the SM charged fermions as well as for {the} active neutrinos, the SM scalar sector has to be enlarged by the inclusion of an extra $SU\left( 2\right) $ scalar doublet $\eta $ as well as two gauge singlet scalars $\xi $ and $\sigma $. These scalars can be expanded as 
\begin{equation}
 \eta =\left( 
\begin{array}{c}
\eta ^{+} \\ 
\frac{1}{\sqrt{2}}\left( \eta _{R}^{0}+i\eta _{I}^{0}\right) 
\end{array}%
\right) , \,\,\, \,\,\,\,
\xi = \frac{1}{\sqrt{2}}\left(\xi_R + i~ \xi_I\right), \,\,\, \,\,\,\, 
\sigma = \frac{1}{\sqrt{2}}\left(v_{\sigma}+\tilde{\sigma}\right).  \label{BSMscalars}
\end{equation}

Furthermore, the SM fermion sector is augmented by adding heavy vector-like up,
down quarks and charged leptons as well as right handed Majorana neutrinos, denoted
as
{\begin{equation}
T_{k}, B_{k}, E_{k}~~(k=1,2) ~~~~~~ \textrm{and}~~~~~~ N_{jR}~~(j=1,2,3),
\end{equation}}
respectively. The complete fermionic and scalar particle contents and their charges under $SU(3)_c \times SU(2)_L \times U(1)_Y \times Z_4$ are given in Tables \ref{fermions} and \ref{scalars}, respectively.

Note that the $vev$ of $\sigma$ will break the $Z_4$ symmetry to a remnant $Z_2$ symmetry. \VKN{The $Z_{2}$ transformation of the fields are defined as 
$f \rightarrow (-1)^{Q_{Z_4}}f$.} Thus, the scalars $\eta$ and $\xi$ are odd under this remnant $Z_2$ symmetry, because of which, these scalars do not acquire $vev$s. Thus,  the preserved $Z_{2}$
symmetry allows for a stable scalar DM candidate, which can be the lightest of the
neutral CP- odd or -even component of $\eta $ or the CP- odd or -even
component of $\xi $. Also, for this reason, we refer to the scalars $\eta$ and $\xi$ as dark scalars from here onwards.

\begin{table}[tp]
\begin{tabular}{|c|c|c|c|c|c|c|c|c|c|c|c|c|c|c|}
\hline
& $q_{nL}$ & $q_{3L}$ & $u_{iR}$ & $d_{iR}$ & $T_{nL}$ & $T_{nR}$ & $B_{nL}$
& $B_{nR}$ & $l_{iL}$ & $e_{nR}$ & $e_{3R}$ & $E_{nL}$ & $E_{nR}$ & $N_{iR}$
\\ \hline
$SU(3)_{C}$ & $\mathbf{3}$ & $\mathbf{3}$ & $\mathbf{3}$ & $\mathbf{3}$ & $%
\mathbf{3}$ & $\mathbf{3}$ & $\mathbf{3}$ & $\mathbf{3}$ & $\mathbf{1}$ & $%
\mathbf{1}$ & $\mathbf{1}$ & $\mathbf{1}$ & $\mathbf{1}$ & $\mathbf{1}$ \\ 
\hline
$SU(2)_{L}$ & $\mathbf{2}$ & $\mathbf{2}$ & $\mathbf{1}$ & $\mathbf{1}$ & $%
\mathbf{1}$ & $\mathbf{1}$ & $\mathbf{1}$ & $\mathbf{1}$ & $\mathbf{2}$ & $%
\mathbf{1}$ & $\mathbf{1}$ & $\mathbf{1}$ & $\mathbf{1}$ & $\mathbf{1}$ \\ 
\hline
$U(1)_{Y}$ & $\frac{1}{6}$ & $\frac{1}{6}$ & $\frac{2}{3}$ & $-\frac{1}{3}$
& $\frac{2}{3}$ & $\frac{2}{3}$ & $-\frac{1}{3}$ & $-\frac{1}{3}$ & $-\frac{1%
}{2}$ & $-1$ & $-1$ & $-1$ & $-1$ & $0$ \\ \hline
$Z_{4}$ & $-2$ & $0$ & $0$ & $0$ & $-1$ & $1$ & $1$ & $-1$ & $0$ & $2$ & $0$
& $-1$ & $1$ & $-1$ \\ \hline
\end{tabular}%
\caption{Fermion charge assignments under the $SU\left( 3\right) _{C}\times
SU\left( 2\right) _{L}\times SU\left( 2\right) _{Y}\times Z_{4}$ symmetry. Here $i=1,2,3$ and $n=1,2$.}
\label{fermions}
\end{table}
\begin{table}[tp]
\begin{tabular}{|c|c|c|c|c|}
\hline
& $\phi$ & $\eta$ & $\xi$ & $\sigma$ \\ \hline
$SU(3)_{C}$ & $\mathbf{1}$ & $\mathbf{1}$ & $\mathbf{1}$ & $\mathbf{1}$ \\ 
\hline
$SU(2)_{L}$ & $\mathbf{2}$ & $\mathbf{2}$ & $\mathbf{1}$ & $\mathbf{1}$ \\ 
\hline
$U(1)_{Y}$ & $\frac{1}{2}$ & $\frac{1}{2}$ & $0$ & $0$ \\ \hline
$Z_{4}$ & $0$ & $-1$ & $1$ & $2$ \\ \hline
\end{tabular}%
\caption{Scalar charge assignments under the $SU\left( 3\right) _{C}\times
SU\left( 2\right) _{L}\times SU\left( 2\right) _{Y}\times Z_{4}$ symmetry.}
\label{scalars}
\end{table}

\begin{figure}[th]
\begin{center}
\includegraphics[width=0.5\textwidth]{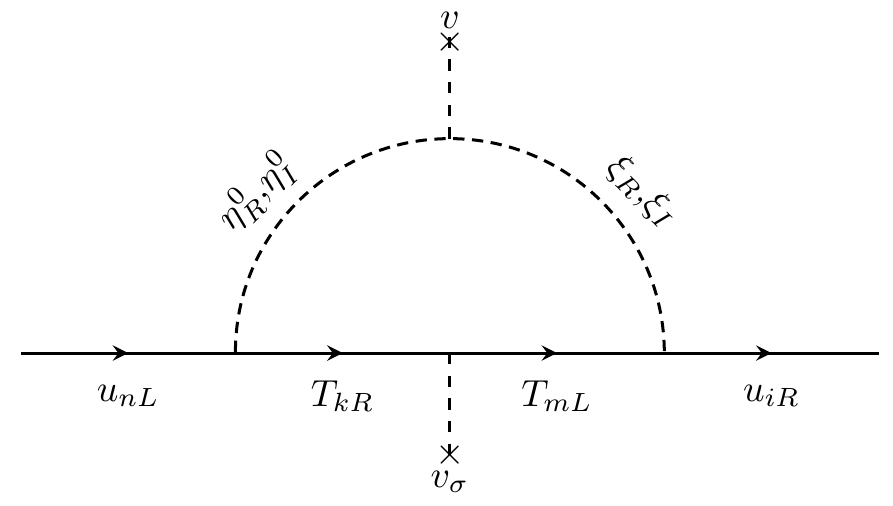}%
\includegraphics[width=0.5\textwidth]{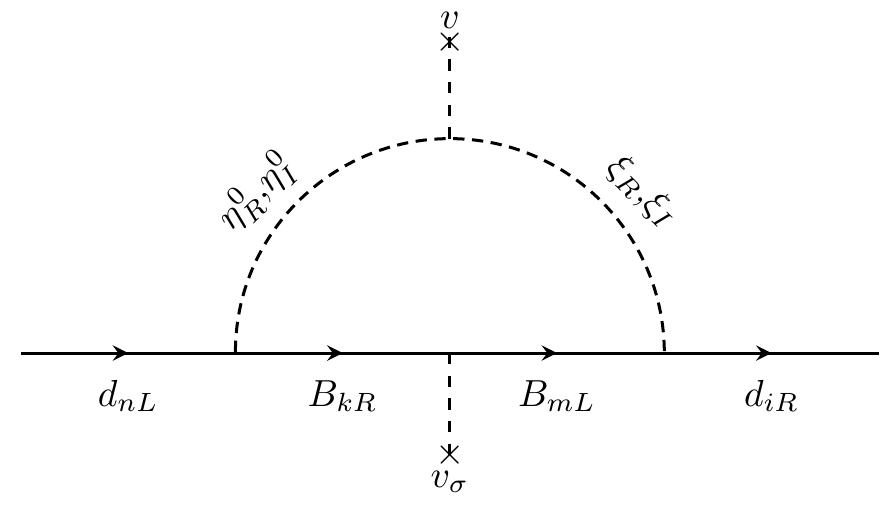}
\end{center}
\caption{Loop Feynman diagrams contributing to the entries of the SM up
(left panel) and SM down (right panel) quark mass matrices. Here $i=1,2,3$
and $k,n,m=1,2$.}
\label{Loopdiagramquarks}
\end{figure}

\begin{figure}[th]
\begin{center}
\includegraphics[width=0.47\textwidth]{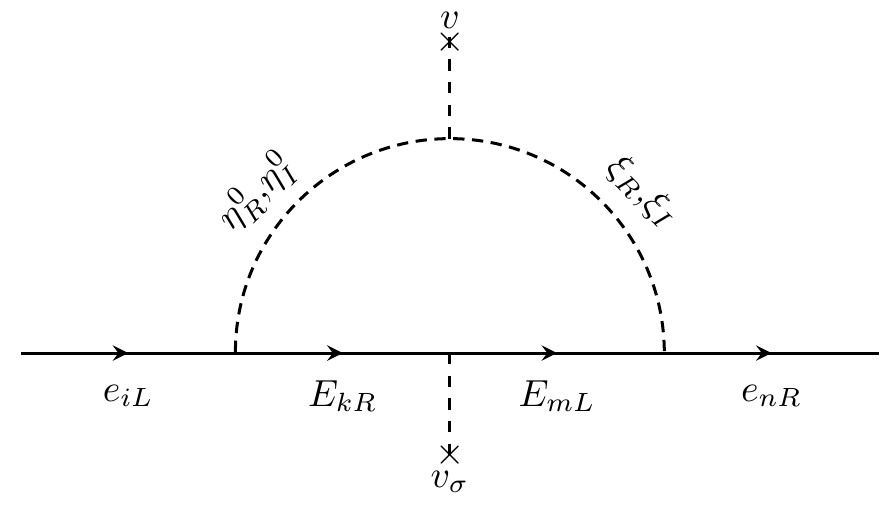} %
\includegraphics[width=0.47\textwidth]{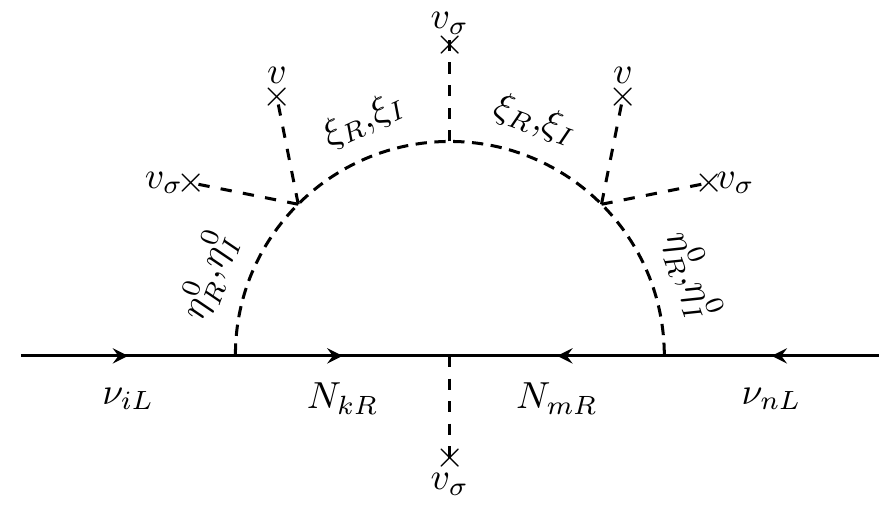}
\end{center}
\caption{Loop Feynman diagrams contributing to the entries of the charged
lepton (left panel) and light active neutrino (right panel) mass matrices.
Here $i,k,m,n=1,2,3$.}
\label{Loopdiagramleptons}
\end{figure}

The complete quark and leptonic Yukawa parts of the Lagrangian that are invariant under the
symmetries of the model are given as
\begin{eqnarray}
-\mathcal{L}_{Y}^{\left( q\right) } &=&\sum_{i=1}^{3}y_{i}^{\left( u\right) }%
\overline{q}_{3L}\widetilde{\phi }u_{iR}+\sum_{i=1}^{3}y_{i}^{\left(
d\right) }\overline{q}_{3L}\phi
d_{iR}+\sum_{n=1}^{2}\sum_{k=1}^{2}y_{nk}^{\left( T\right) }\overline{q}_{nL}%
\widetilde{\eta }T_{kR}+\sum_{k=1}^{2}\sum_{i=1}^{3}x_{ki}^{\left( u\right) }%
\overline{T}_{kL}\xi ^{\ast }u_{iR}  \notag \\
&&+\sum_{n=1}^{2}\sum_{k=1}^{2}y_{nk}^{\left( B\right) }\overline{q}%
_{nL}\eta B_{kR}+\sum_{k=1}^{2}\sum_{n=1}^{2}x_{kn}^{\left( d\right) }%
\overline{B}_{kL}\xi d_{nR}  \notag \\
&&+\sum_{k=1}^{2}\sum_{n=1}^{2}\left( z_{T}\right) _{kn}\overline{T}%
_{kL}\sigma ^{\ast }T_{nR}+\sum_{k=1}^{2}\sum_{n=1}^{2}\left( z_{B}\right)
_{kn}\overline{B}_{kL}\sigma B_{nR}+H.c,  \label{Lyq}
\end{eqnarray}%
and,
\begin{eqnarray}
-\mathcal{L}_{Y}^{\left( l\right) } &=&\sum_{i=1}^{3}y_{i}^{\left(e\right) }%
\overline{l}_{iL}\phi e_{3R}+\sum_{i=1}^{3}\sum_{k=1}^{2}y_{ik}^{\left(
E\right) }\overline{l}_{iL}\eta
E_{kR}+\sum_{k=1}^{2}\sum_{n=1}^{2}x_{kn}^{\left( e\right) }\overline{E}%
_{kL}\xi e_{nR}+\sum_{i=1}^{3}{\sum_{j=1}^{3}y_{ij}^{\left( N\right) }%
\overline{l}_{iL}\widetilde{\eta }N_{jR}}  \notag \\
&&+\sum_{k=1}^{2}\sum_{n=1}^{2}\left( z_{E}\right) _{kn}\overline{E}%
_{kL}\sigma E_{nR}+{\sum_{i=1}^{3}\sum_{j=1}^{3}\left( z_{N}\right)
_{ij}N_{iR}\overline{N_{jR}^{C}}\sigma }+H.c ,\label{Lyl}
\end{eqnarray}%
respectively. The complete scalar potential and the masses and mixing of the scalars are discussed in the Appendix. From the Lagrangian,
the first two generations of the SM charged fermions as well as
the three generations of light active neutrinos obtain their masses from one loop level
radiative seesaw mechanism as shown in the Feynman diagrams of Figures \ref{Loopdiagramquarks} and  \ref{Loopdiagramleptons}. 
%{\red \bf{(Note : The expressions for charged fermion masses are not given in the paper.)}}. 
{ Note that the dark scalars $\eta$ and $\xi$ run in these loop diagrams and thereby bring in a connection between the DM, active $\nu$ mass generation and fermion mass hierarchy. Thus our model acts as a generalized version of the combination of scotogenic model~\cite{Tao:1996vb} and IDM \cite{Deshpande:1977rw}. }

%%%%%%%%%%%%%%%%%%%%%%%%%%%
%%%%%%%%%%%%%%%%%%%%%%%%%%%%%%
%%%%%%%%%%%%%%%%%%%%%%%%%%%
%%%%%%%%%%%%%%%%%%%%%%%%%%%%%%
%%%%%%%%%%%%%%%%%%%%%%%%%%%
%%%%%%%%%%%%%%%%%%%%%%%%%%%%%%

%From the loop diagrams of Figures \ref{Loopdiagramquarks} and \ref{Loopdiagramleptons}
From the charged fermion Yukawa interactions and evaluating the diagrams in Figs.\ref{Loopdiagramquarks} and \ref{Loopdiagramleptons}, we find that the entries of the mass matrices for SM charged fermions are given as
\begin{eqnarray}
\left( M_{U}\right) _{ni} &=&\dsum\limits_{k=1}^{2}\frac{y_{nk}^{\left(
T\right) }x_{ki}^{\left( u\right) }m_{T_{k}}}{16\pi ^{2}}\left\{ \left[
f\left( m_{S_{1}}^{2},m_{T_{k}}^{2}\right) -f\left(
m_{S_{2}}^{2},m_{T_{k}}^{2}\right) \right] \sin 2\theta _{S}-\left[ f\left(
m_{A_{1}}^{2},m_{T_{k}}^{2}\right) -f\left(
m_{A_{2}}^{2},m_{T_{k}}^{2}\right) \right] \sin 2\theta _{A}\right\} , 
\notag \\
\left( M_{U}\right) _{3i} &=&y_{i}^{\left( u\right) }\frac{v}{\sqrt{2}}%
,\qquad \qquad k,n=1,2,\qquad \qquad i=1,2,3.
\end{eqnarray}
\begin{eqnarray}
\left( M_{D}\right) _{ni} &=&\dsum\limits_{k=1}^{2}\frac{y_{nk}^{\left(
B\right) }x_{kn}^{\left( d\right) }m_{B_{k}}}{16\pi ^{2}}\left\{ \left[
f\left( m_{S_{1}}^{2},m_{B_{k}}^{2}\right) -f\left(
m_{S_{2}}^{2},m_{B_{k}}^{2}\right) \right] \sin 2\theta _{S}-\left[ f\left(
m_{A_{1}}^{2},m_{B_{k}}^{2}\right) -f\left(
m_{A_{2}}^{2},m_{B_{k}}^{2}\right) \right] \sin 2\theta _{A}\right\} , 
\notag \\
\left( M_{D}\right) _{3i} &=&y_{i}^{\left( d\right) }\frac{v}{\sqrt{2}}%
,\qquad \qquad k,n=1,2,\qquad \qquad i=1,2,3.
\end{eqnarray}
\begin{eqnarray}
\left( M_{l}\right) _{ni} &=&\dsum\limits_{k=1}^{2}\frac{y_{nk}^{\left(
E\right) }x_{kn}^{\left( e\right) }m_{E_{k}}}{16\pi ^{2}}\left\{ \left[
f\left( m_{S_{1}}^{2},m_{E_{k}}^{2}\right) -f\left(
m_{S_{2}}^{2},m_{E_{k}}^{2}\right) \right] \sin 2\theta _{S}-\left[ f\left(
m_{A_{1}}^{2},m_{E_{k}}^{2}\right) -f\left(
m_{A_{2}}^{2},m_{E_{k}}^{2}\right) \right] \sin 2\theta _{A}\right\} , 
\notag \\
\left( M_{l}\right) _{3i} &=&y_{i}^{\left( e\right) }\frac{v}{\sqrt{2}}%
,\qquad \qquad k,n=1,2,\qquad \qquad i=1,2,3.
\end{eqnarray}

\VKN{In the above equations, the function $f(m_1^2,m_2^2)$ is defined as
\be f(m_1^2, m_2^2) = \frac{m_1^2}{m_1^2 - m_2^2}~\textrm{ln}~\frac{m_1^2}{m_2^2}, \ee } and the fields $S_{1,2}$ and $A_{1,2}$ 
correspond to the physical dark CP even and CP odd neutral scalar mass eigenstates, respectively. As shown in detail in the Appendix, they are related to the gauge eigenstates as
\begin{equation}
\left( 
\begin{array}{c}
S_{1} \\ 
S_{2}%
\end{array}%
\right) = (R_S)^T \left( \begin{array}{c}
\eta _{R} \\ 
\xi _{R}%
\end{array}%
\right)
=
\left( 
\begin{array}{cc}
\cos \theta _{S} & \sin \theta _{S} \\ 
-\sin \theta _{S} & \cos \theta _{S}%
\end{array}%
\right) \left( 
\begin{array}{c}
\eta _{R} \\ 
\xi _{R}%
\end{array}%
\right) , \hspace{0.5cm}\left( 
\begin{array}{c}
A_{1} \\ 
A_{2}%
\end{array}%
\right) = (R_A)^T \left( \begin{array}{c}
\eta _{I} \\ 
\xi _{I}%
\end{array}%
\right) =
\left( 
\begin{array}{cc}
\cos \theta _{A} & \sin \theta _{A} \\ 
-\sin \theta _{A} & \cos \theta _{A}%
\end{array}%
\right) \left( 
\begin{array}{c}
\eta _{I} \\ 
\xi _{I}%
\end{array}%
\right).
\end{equation}
Note that the expressions for the charged fermion masses depend on the above scalar mixing angles due to the presence of the interaction vertex $C_{1}\left( \phi ^{\dagger }\eta \xi \right) $  in the loop diagrams for the masses of the charged fermions. The expressions for these mixing angles depend on the parameters in the potential and are given in the Appendix. Also, evaluating the diagram in the left hand side of Fig.\ref{Loopdiagramleptons} results in the following expression for the active neutrino mass matrix:
{%
\begin{eqnarray}
\left( M_{\nu }\right) _{ij} &=& \sum_{\alpha =1}^{2}{ \sum_{k=1}^{3}}\frac{%
\left( y^{\left( N\right) }R_{\widetilde{N}}\right) _{ik}\left( y^{\left(
N\right) }R_{\widetilde{N}}\right) _{jk}m_{\widetilde{N}_{k}}}{16\pi ^{2}}%
\left[ \left( \left( R_{S}\right) _{\alpha 1}\right) ^{2}\frac{%
m_{_{S_{\alpha }}}^{2}}{m_{_{S_{\alpha }}}^{2}-m_{\widetilde{N}_{k}}^{2}}\ln
\left( \frac{m_{_{S_{\alpha }}}^{2}}{m_{\widetilde{N}_{k}}^{2}}\right)
\right.  \notag \\
&-&\left. \left( \left( R_{A}\right) _{\alpha 1}\right) ^{2}\frac{%
m_{_{A_{\alpha }}}^{2}}{m_{A}^{2}-m_{\widetilde{N}_{k}}^{2}}\ln \left( \frac{%
m_{A_{\alpha }}^{2}}{m_{\widetilde{N}_{k}}^{2}}\right) \right] ,\hspace{1cm}%
i,j=1,2,3{\small ,}\label{numasseqn}
\end{eqnarray}%
{where $R_{\widetilde{N}}$, diagonalizes the sterile
neutrino mass matrix $m_{\widetilde{N}}$ and $m_{\widetilde{N}_k}$ are the corresponding eigenvalues. 
Again, note the dependence on the scalar mixing due to the presence of the vertex  $C_{2} \left( \xi ^{2}\sigma \right)$ in the loop diagram for neutrino mass.}

To summarize, our proposed model corresponds to an extension of the IDM theory, where the
scalar sector is augmented by the inclusion of two gauge singlet scalars
whereas the fermion sector is increased by some charged vector-like
fermions and right handed Majorana neutrinos. The SM gauge symmetry is
supplemented by the inclusion of the $Z_{4}$ discrete symmetry, whose
spontaneous breaking down to a preserved $Z_{2}$ symmetry yields one loop
level masses for the first and second families of SM charged fermions as
well as for the active light Majorana neutrinos. It is worth mentioning that the
remnant preserved $Z_{2}$ symmetry prevents tree level masses for the first and
second generation of SM charged fermions as well as for the active
neutrinos, thereby guaranteeing the radiative nature of the seesaw mechanisms that
produce these masses.

{\subsection{Modified Casas-Ibarra Parametrization for the neutrino Yukawa coupling}

In this section, we discuss a simple parametrization for the neutrino Yukawa coupling matrix $y^{(N)}$, inspired by the Casas-Ibarra (CI) parametrization proposed in \cite{Casas:2001sr}  for the canonical type-I seesaw mechanism. The idea is to express the Yukawa coupling matrix in terms of the PMNS neutrino mixing matrix, active light neutrino masses and the masses of the heavy particles that contribute to light neutrino mass generation. By doing this, we ensure that the Yukawa coupling matrix that we take always reproduces the correct active light neutrino masses and mixing.

For simplicity, we take the sterile
neutrino mass matrix $m_{\widetilde{N}}$ to be diagonal so that $R_{\widetilde{N}}$ becomes the identity matrix} and denoting
\begin{equation}
\Lambda_k = \sum_{\alpha =1}^{2}\frac{1}{16\pi ^{2}}
\Big[ \left( \left( R_{S}\right) _{\alpha 1}\right) ^{2}\frac{m_{_{S_{\alpha }}}^{2}}{m_{_{S_{\alpha }}}^{2}-m_{\widetilde{N}_{k}}^{2}}\textrm{ln}
\left( \frac{m_{_{S_{\alpha }}}^{2}}
{m_{\widetilde{N}_{k}}^{2}}\right) -\left( \left( R_{A}\right)_{\alpha 1}\right) ^{2}\frac{m_{_{A_{\alpha }}}^{2}}
{m_{A}^{2}-m_{\widetilde{N}_{k}}^{2}}\textrm{ln} \left( 
\frac{m_{A_{\alpha }}^{2}}{m_{\widetilde{N}_{k}}^{2}}\right) \Big],
\end{equation}%
the light neutrino mass matrix in Eqn.\ref{numasseqn} can be written as
\begin{equation}
 M_{\nu }  = y^{(N)} \Lambda {y^{(N)}}^T.
\end{equation}
This equation can be inverted to express the Yukawa coupling matrix in terms of the neutrino oscillation parameters using the modified CI parametrization as~\cite{Casas:2001sr}
\begin{equation}
y^{(N)} = U^*\sqrt{M_\nu^{diag}} R \sqrt{\Lambda_k^{-1}} . \label{CI}
\end{equation}
In the above equation, ${M_\nu}^{diag}$ is the diagonal light neutrino mass matrix, $U$ is the PMNS mixing matrix and $R$ is a general $3 \times 3$ complex orthogonal matrix parametrized by the three complex mixing angles, $z_1$, $z_2$ and $z_3$.

\section{Phenomenological Implications}
\label{pheno}

In this section, we discuss the various phenomenological implications of our model. In particular, we study the predictions for charged lepton flavor violation, muon and electron magnetic moments, bounds from electroweak precision observables, collider constraints, leptogenesis and dark matter phenomenology.

\subsection{Lepton flavor violation}\label{seclfv}

\begin{figure}[tbp]
\includegraphics[width=0.5\textwidth]{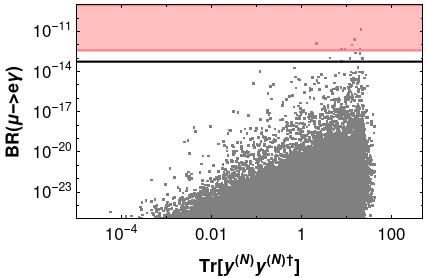}
\caption{{\VKN{Correlation of BR $(\mu \rightarrow e \gamma)$ against Tr$[{y^{(N)}}^\dag y^{(N)} ]$ in our model. All the displayed points satisfy the perturbativity bounds on $y^{(N)}$.} The pink shaded region is disfavored by the existing bound from MEG~\cite{MEG:2016leq} whereas the black horizontal line corresponds to  the projected future sensitivity of MEG-II~\cite{Baldini:2013ke,Meucci:2022qbh}. In generating this figure, we have used the modified CI parametrization for $y^{(N)}$ as given in Eqn.~\ref{CI}.}}
\label{lfv2}
\end{figure}

One of the interesting features of the model that we are studying is the predictions for charged lepton flavour violating (LFV) decays, such as $\mu \rightarrow e \gamma$, which are strongly constrained by experiments. These decays occur at the one-loop level and are mediated by the exchanges of the neutral fermions and the charged scalars. The branching fraction for the radiative two-body decay process $\ell_i \rightarrow \ell_j \gamma$, where $i = \mu,\tau$ is given as~\cite{Ma:2001mr,Toma:2013zsa,Vicente:2014wga,Lindner:2016bgg}
	\begin{equation}\label{eq:mu_to_egamma}
	\mathrm{BR}\left( l_{i}\rightarrow l_{j}\gamma \right) =\frac{3\left( 4\pi \right)
		^{3}\alpha _{\mathrm{em}}}{4G_{F}^{2}}\left\vert \frac{\sum_{k=1}^{3}(y^{(N)})_{ks}\left( V_{lL}^{\dagger }\right) _{ik}  \sum_{k,m=1}^{3}(y^{(N)})_{ms}\left( V_{lL}^{\dagger }\right) _{jm}}{2\left( 4\pi \right) ^{2}m_{\eta
			^{\pm }}^{2}}F\left( \frac{m_{\tilde{N}_{k}}^{2}}{m_{\eta ^{\pm }}^{2}}\right)
	\right\vert ^{2}\mathrm{BR}\left( l_{i}\rightarrow l_{j}\nu _{i}\overline{\nu _{j}}%
	\right) \,,  
	\end{equation}
	with $s=1,2,3$. Here, $\alpha_{\mathrm{em}}$ is the fine-structure constant, $G_F$ the Fermi constant, $V$ is the left-handed charged lepton mixing matrix and $m_{\eta ^{\pm }}$ is the mass of the charged scalar components of the $\mathrm{SU(2)_{L}}$ inert doublet $\eta $. The loop function $F$ is given as
 \begin{equation}\label{eq:loop_function}
	F\left( x\right) =\frac{1-6x+3x^{2}+2x^{3}-6x^{2}\ln x}{6\left( 1-x\right)}.
	\end{equation}
	The most stringent bounds for LFV come from muon decay measurements, namely {$\mu\rightarrow e \gamma$}. The MEG experiment puts an upper bound on the branching ratio as comes $\mathrm{BR}\left(\mu \rightarrow e\gamma \right) < 4.2 \times 10^{-13}$ \cite{MEG:2016leq}. 

In Fig.\ref{lfv2}, we show the predictions for $\mathrm{BR}\left(\mu \rightarrow e\gamma \right)$ in our model as a function of Tr$[{ y^{(N)} y^{(N)}}^\dag ]$. The pink shaded region is disfavored by the existing bound from MEG whereas the black horizontal line corresponds to  the projected future sensitivity of $6 \times 10^{-14}$ for MEG-II~\cite{Baldini:2013ke,Meucci:2022qbh}. \VKN{All the points that are displayed are allowed by the perturbativity bounds on $y^{(N)}$, $y^{(N)}_{ij} < \sqrt{4 \pi}$.} In generating this figure, we have used the modified CI parametrization for $y^{(N)}$ as given in Eqn.~\ref{CI}. {The masses of the scalars $\eta^\pm$, $S_1$, $S_2$, $A_1$ and $A_2$ are all varied in the range $[300~\textrm{GeV}, 1.7~\textrm{TeV}]$, the masses of the lightest heavy neutrino, $m_{\tilde{N}_1}$ is varied in the range $[200~\textrm{GeV}, 200~\textrm{TeV}]$,
$ m_{\tilde{N}_2}$ and $m_{\tilde{N}_3}$ are varied in ranges $[m_{\tilde{N}_1}, 2000~\textrm{TeV}]$, the scalar mixing angles entering the expression for $\Lambda_k$ are varied in the range $[0,1]$, the active light neutrino mixing angles, mass-squared differences and CP phase are varied in the $3\sigma$ ranges~\cite{deSalas:2020pgw} and the active neutrino Majorana phases are varied in the range $[0,\pi]$. For the orthogonal matrix $R$ in Eqn.~\ref{CI}, the complex mixing angles $z_1$ and $z_2$ are taken to be $0$ whereas $z_3$ is taken as $x_3 - I x_3$ with $x_3$ varying in the range $[0,50]$.}

From this figure, it is evident that the majority of the predicted values lie with the current limits. However, it is noteworthy that a small number of data points corresponding to significantly large Yukawa couplings are disfavored by MEG whereas a few points are present within the future discovery sensitivity of MEG-II.

\subsection{Muon and electron anomalous magnetic moments}

%%%%%%%%%%%%%%%%%%%%%%%%%%%%%%%%%%%%%%%%%%%%%%%%%%%%%%%%%%%%%%%%%%%%%%%%%%
In this subsection we analyze the implications of our model for the muon
and the electron anomalous magnetic moments. The contributions to these mainly arise from the one-loop
diagrams involving the exchange of electrically neutral scalars and charged vector like leptons running in the internal lines of the loop as shown in Figure \ref{fig:gminus2diagram}.

%%%%%%%%%%%%%%%%%%%%%%%%%%%%%%
\begin{figure}[th]
\begin{center}
\includegraphics[width=0.45\textwidth]{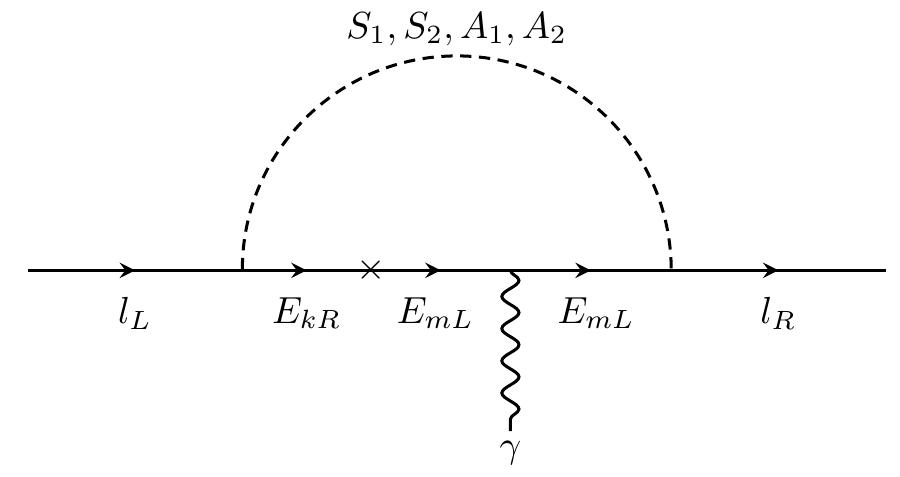} %
\includegraphics[width=0.45\textwidth]{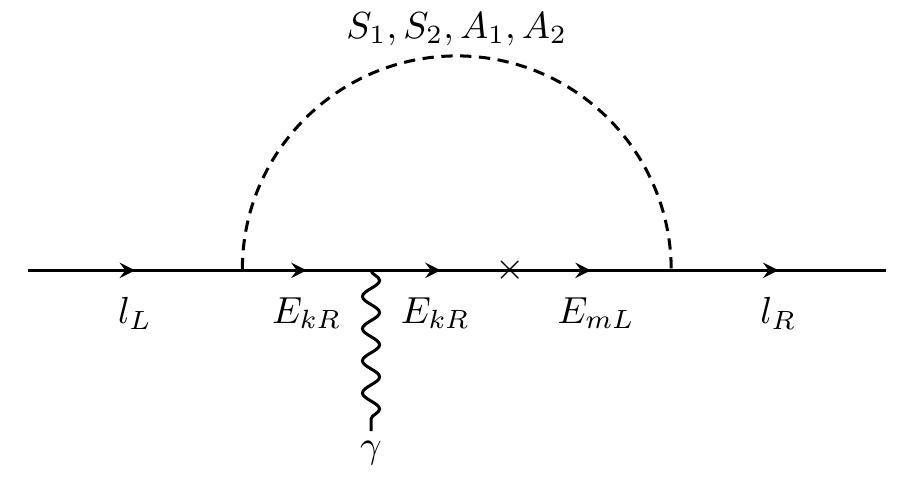}
\end{center}
\caption{Feynman diagrams contributing to the anomalous magnetic moment of the electron and muon.}
\label{fig:gminus2diagram}
\end{figure}
%%%%%%%%%%%%%%%%%%%%%%%%%%%

\VKN{Following the results from \cite{Crivellin:2018qmi}, the leading
contributions to the muon and electron anomalous magnetic moments in our model take the form:
\be a_\mu = - \frac{4 m_\mu}{e}~\textrm{Re}\big[ c^{22}\big]~~~~~~~, ~~~~~~~~a_e = - \frac{4 m_e}{e}~\textrm{Re}\big[ c^{11}\big], \ee
where the Wilson coefficient $c^{fi}$ is given as,
\begin{eqnarray}
c^{fi} = \frac{e}{32 \pi^2} \sum_{k,p = 1}^{2} &\Bigg(& \Big( {(\beta_{kf})}^* \gamma_{ki} {(R_S)}_{1p} {(R_S)}_{2p} m_{E_k} I_1(m_{E_k},m_{S_p}) - {(\beta_{kf})}^* \gamma_{ki} {(R_A)}_{1p} {(R_A)}_{2p} m_{E_k} I_1(m_{E_k},m_{A_p}) \Big) 
\notag \\
&&+  \Big(  
m_{l_i} {(\beta_{kf})}^* \beta_{ki} {(R_S)}_{1p}^2 + 
m_{l_f} {(\gamma_{kf})}^* \gamma_{ki} {(R_S)}_{2p}^2
\Big) I_2(m_{E_k},m_{S_p}) \\
&&+ \Big(  
m_{l_i} {(\beta_{kf})}^* \beta_{ki} {(R_A)}_{1p}^2 + 
m_{l_f} {(\gamma_{kf})}^* \gamma_{ki} {(R_A)}_{2p}^2
\Big) I_2(m_{E_k},m_{A_p}) \Bigg)\notag.
\end{eqnarray}

In the above equation, the loop factors $I_{1}(m_1, m_2)$ and $I_{2}(m_1, m_2)$ have the form:
\begin{equation}
I_{1}(m_1, m_2) = \frac{f_1\Big(\frac{m_1^2}{m_2^2}\Big) - g_1\Big(\frac{m_1^2}{m_2^2}\Big)}{m_2^2} ~~~~,~~~~~ I_{2}(m_1, m_2) = \frac{f_2\Big(\frac{m_1^2}{m_2^2}\Big) - g_2\Big(\frac{m_1^2}{m_2^2}\Big)}{m_2^2},
\end{equation}%
where, the functions $f_1$, $f_2$, $g_1$, and $g_2$ are given as,
\be f_1(x) = \frac{x^2 - 1 - 2 x~\textrm{log}(x)}{4(x-1)^3} = 2 g_2(x),~~  f_2(x) = \frac{2 x^3 + 3x^2 - 6x + 1 - 6 x^2~\textrm{log}(x)}{24(x-1)^4},  ~~  g_1(x) = \frac{x-1-\textrm{log}(x)}{2(x-1)^2}. \ee

The dimensionless parameters $\beta _{ik}$ and $\gamma _{ik}$ are given by:
\be
\beta_{ik} = ({y^{(E)}}^\dag V_{lL}^\dag )_{ik}~~~~~, ~~~~~ \gamma_{ik} = (x^{(e)} V_{lR}^\dag )_{ik},
\ee
where $V_{lL}$ and $V_{lR}$ are the rotation matrices that diagonalize the charged lepton matrix $M_l$ according to the relation
\begin{equation}
V_{lL}^{\dagger } {M}_l V_{lR}=diag\left( m_{e},m_{\mu },m_{\tau
}\right).
\end{equation}
The muon and electron anomalous magnetic moments are
constrained to be in the ranges \cite{Abi:2021gix,Morel:2020dww,Muong-2:2023cdq},
\begin{eqnarray}
\left( \Delta a_{\mu }\right) _{\exp } &=&\left( 2.49\pm 0.48\right) \times
10^{-9},  \notag \\
(\Delta a_{e})_{\text{exp}} &=&(4.8\pm 3.0)\times 10^{-13}.
\end{eqnarray}%

In addition to the anomalous magnetic moments, the diagrams in Figure \ref{fig:gminus2diagram} can also contribute to the electric dipole moments (EDM) of the electron and muon. These are given as,
\be  d_e = -2\textrm{Im}[c^{11}]~~~~~~~,~~~~~~ d_\mu =  -2\textrm{Im}[c^{22}]. \ee
The EDM of electron is very strongly constrained by experiments, the most stringent bound being \cite{Roussy:2022cmp},
\be d_e < 4.1 \times 10^{-30} e.cm .\label{eEDM} \ee

Figure \ref{gminus2c} shows the allowed parameter space in the $m_{E}-m_{S_{1}}$  plane consistent with the muon and electron anomalous magnetic moments. In plotting this figure, we have varied the masses of the scalar $S_1$ in the range 200 GeV - 2 TeV whereas the masses of the other three scalars $S_2$, $A_1$, and $A_2$ are varied in the range $m_{S_1} \pm 1$ GeV. The two exotic fermions $E_1$ and $E_2$ are taken to be degenerate with their masses varying in the range 100 GeV - 2 TeV. Further, the effective couplings $\beta$ and $\gamma$ are taken as the input parameters, with $\beta_{12}=\beta_{21} = \gamma_{12}=\gamma_{21} = 0$, whereas $|\beta_{11}|,~ \beta_{21}, ~ \gamma_{12}, $ and $\gamma_{21}$ are varied in the range $0-\sqrt{4\pi}$. Only $\beta_{11}$ is taken to be complex with its phase varying in the range $0-2\pi$. All the points displayed in Figure \ref{gminus2c} also satisfy the bound on the electron EDM given by Eq.~\ref{eEDM}. In fact, this can be made sure in our analysis by taking very small values for the phase of $\beta_{11}$ since $d_E$ depends on the imaginary part of the Wilson coefficient $c^{11}$, whereas the magnetic moments depend on the real part of $c^{ii}~(i=1,2)$. We note that the values of scalar masses beyond $\sim 1$ TeV and that of the exotic leptons beyond $\sim 800$ GeV are disfavored by the constraints on the anomalous magnetic moments.}

\begin{figure}[tbp]
\centering
\includegraphics[width=0.45\textwidth]{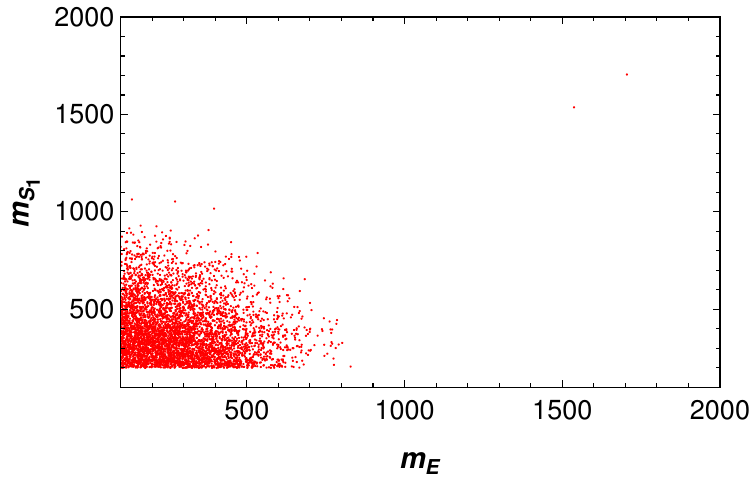}
\caption{{Parameter space in the $m_{E}-m_{S_{1}}$ plane that is consistent with the measured values of the electron and muon anomalous magnetic moments.}}
\label{gminus2c}
\end{figure}

\subsection{Oblique parameters}

%%%%%%%%%%%%%%%%%%%%%%%%%%%%%%%%%%%%%%%%%%%%%%%%%%%%%%%%%
The extra scalars in our model, in particular, the inert scalar
doublet, affect the oblique corrections of the SM which are parameterized
in terms of the well-known quantities $T$, $S$ and $U$, defined as~\cite%
{Altarelli:1990zd, Peskin:1991sw, Barbieri:2004qk} 
\begin{align}
T& =\left. \frac{\Pi _{33}\left( q^{2}\right) -\Pi _{11}\left( q^{2}\right) 
}{\alpha _{\text{EM}}(M_{Z})\,M_{W}^{2}}\right\vert _{q^{2}=0}, \\
S& =\left. \frac{2\,\sin 2\theta _{W}}{\alpha _{\text{EM}}(M_{Z})}\frac{d\Pi
_{30}\left( q^{2}\right) }{dq^{2}}\right\vert _{q^{2}=0}, \\
U& =\left. \frac{4\sin ^{2}\theta _{W}}{\alpha _{\text{EM}}(M_{Z})}\left( 
\frac{d\Pi _{33}\left( q^{2}\right) }{dq^{2}}-\frac{d\Pi _{11}\left(
q^{2}\right) }{dq^{2}}\right) \right\vert _{q^{2}=0}.
\end{align}%
Here $\Pi _{11}\left( 0\right) $, $\Pi _{33}\left( 0\right) $, and $\Pi
_{30}\left( q^{2}\right) $ are the vacuum polarization amplitudes with the $%
\{W_{\mu }^{1}\,,W_{\mu }^{1}\}$, $\{W_{\mu }^{3},\,W_{\mu }^{3}\}$ and $%
\{W_{\mu }^{3},\,B_{\mu }\}$ external gauge bosons, respectively. We note
that in the definitions of the parameters $S$, $T$ and $U$, the new physics
scale is assumed to be heavy compared to $M_{W}$ and $M_{Z}$. As their
values are measured in high-precision experiments, they act as a constraint
on the validity of our model. In this section, we determine the one-loop
contributions to $S$, $T$, $U$ in our model and find the parameter space
where the oblique parameter constraints can be successfully accommodated.

The one-loop contributions to the oblique parameters arising from the inert
scalar exchange are given by: 
\begin{align}
T& \simeq \frac{1}{16\pi ^{2}\,v^{2}\,\alpha _{\text{EM}}(M_{Z})}\left[
\sum_{i=1}^{2}\sum_{j=1}^{2}\left( \left( R_{S}\right) _{1i}\right)
^{2}\left( \left( R_{A}\right) _{1j}\right) ^{2}\,F\left(
m_{S_{i}^{0}}^{2},\,m_{A_{j}^{0}}^{2}\right) +m_{\eta ^{\pm }}^{2}\right. 
\notag  \label{eq:stu} \\
& \qquad -\left. \sum_{i=1}^{2}\left( \left( R_{S}\right) _{1i}\right)
^{2}\,F\left( m_{S_{i}^{0}}^{2},\,m_{\eta ^{\pm }}^{2}\right)
-\sum_{i=1}^{2}\left( \left( R_{A}\right) _{1i}\right) ^{2}\,F\left(
m_{A_{i}^{0}}^{2},\,m_{\eta ^{\pm }}^{2}\right) \right] , \\
S& \simeq \sum_{i=1}^{2}\sum_{j=1}^{2}\frac{\left( \left( R_{S}\right)
_{1i}\right) ^{2}\left( \left( R_{A}\right) _{1j}\right) ^{2}}{12\pi }%
\,K\left( m_{S_{i}^{0}}^{2},\,m_{A_{j}^{0}}^{2},\,m_{\eta ^{\pm
}}^{2}\right) , \\
U& \simeq -S+\sum_{i=1}^{2}\left[ \left( \left( R_{A}\right) _{1i}\right)
^{2}\,K_{2}\left( m_{A_{j}^{0}}^{2},\,m_{\eta ^{\pm }}^{2}\right) +\left(
\left( R_{S}\right) _{1i}\right) ^{2}\,K_{2}\left(
m_{S_{i}^{0}}^{2},\,m_{\eta ^{\pm }}^{2}\right) \right] ,
\end{align}%
where we introduce the functions~\cite{CarcamoHernandez:2015smi}
\begin{align}
F\left( m_{1}^{2},\,m_{2}^{2}\right) & =\frac{m_{1}^{2}\,m_{2}^{2}}{%
m_{1}^{2}-m_{2}^{2}}\,\ln \left( \frac{m_{1}^{2}}{m_{2}^{2}}\right) , \\
K\left( m_{1}^{2},\,m_{2}^{2},\,m_{3}^{2}\right) & =\frac{1}{\left(
m_{2}^{2}-m_{1}^{2}\right) ^{3}}\left[ m_{1}^{4}\left(
3\,m_{2}^{2}-m_{1}^{2}\right) \,\ln \left( \frac{m_{1}^{2}}{m_{3}^{2}}%
\right) -m_{2}^{4}\left( 3\,m_{1}^{2}-m_{2}^{2}\right) \ln \left( \frac{%
m_{2}^{2}}{m_{3}^{2}}\right) \right.  \notag \\
& \qquad -\left. \frac{1}{6}\left[ 27\,m_{1}^{2}m_{2}^{2}\left(
m_{1}^{2}-m_{2}^{2}\right) +5\left( m_{2}^{6}-m_{1}^{6}\right) \right] %
\right], \\
K_2(m_1^2,m_2^2)&=\frac{%
-5m_{1}^{6}+27m_{1}^{4}m_{2}^{2}-27m_{1}^{2}m_{2}^{4}+6\left(
m_{1}^{6}-3m_{1}^{4}m_{2}^{2}\right) \ln \left( \frac{m_{1}^{2}}{m_{2}^{2}}%
\right) +5m_{2}^{6}}{6\left( m_{1}^{2}-m_{2}^{2}\right) ^{3}}
\end{align}%

The above given functions have the following properties:
\begin{align}
& \lim_{m_{2}\rightarrow m_{1}}F\left( m_{1}^{2},\,m_{2}^{2}\right)
=m_{1}^{2}\,, \\
& \lim_{m_{1}\rightarrow
m_{3}}K(m_{1}^{2},\,m_{2}^{2},\,m_{3}^{2})=K_{2}(m_{2}^{2},\,m_{3}^{2})\,, \\
& \lim_{m_{1}\rightarrow
m_{2}}K(m_{1}^{2},\,m_{2}^{2},\,m_{3}^{2})=K_{1}(m_{2}^{2},\,m_{3}^{2})=\ln
\left( \frac{m_{2}^{2}}{m_{3}^{2}}\right) , \\
& \lim_{m_{2}\rightarrow
m_{3}}K(m_{1}^{2},\,m_{2}^{2},\,m_{3}^{2})=K_{2}(m_{1}^{2},\,m_{3}^{2}).
\end{align}
It is worth mentioning that, from the properties of the loop functions appearing in the expressions for the oblique $S$, $T$ and $U$ parameters given in \cite{Grimus:2007if,Grimus:2008nb,CarcamoHernandez:2015smi}, {it follows that the contributions to these parameters arising from new physics will vanish in the limit of degenerate heavy BSM scalars, in the multiHiggs doublet models.} Thus, in multiHiggs doublet models, a spectrum of the {BSM} scalars with a moderate mass splitting will be favoured by electroweak precision tests. The values allowed for $S$, $T$, $U$ from the PDG electroweak fit~ \cite%
{Zyla:2020zbs} , $S=-0.01\pm 0.10,T=0.03\pm 0.12$ and $U=0.02\pm 0.11$, are
in good agreement with the SM prediction $S=T=U=0$.

\begin{figure}[tbp]
\centering
\includegraphics[width=9.5cm, height=7.5cm]{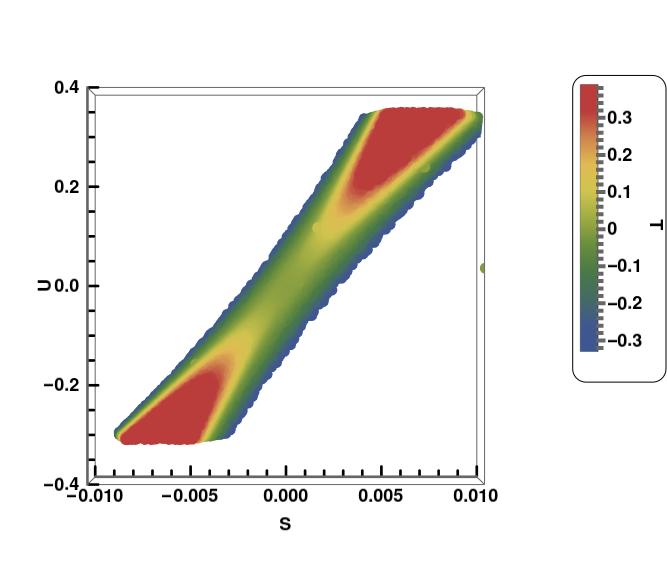}
\includegraphics[width=7.5cm,height=6.5cm]{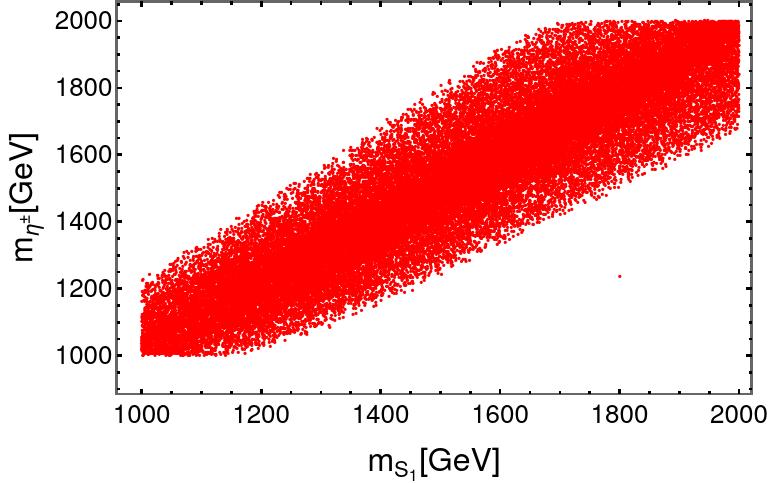} 
\caption{ {Correlations of the $S, T$ and $U$ parameters (left panel) and correlation of the masses of the charged scalar, $\eta^\pm$ against that of the lightest CP even scalar, $S_1$, for the points that satisfy the constraints on $S,T$ and $U$ parameters (right panel). In plotting these figures, the masses of the scalars $A_1,A_2,S_1,S_2$ and $\eta^\pm$ are varied in the range $[1,2]$ TeV.}}
\label{figSTU}
\end{figure}

In the left panel of Fig.~\ref{figSTU}, we have shown the predictions for correlations of the $S, T$ and $U$ parameters in our model. The $S$ and $U$ parameters are shown in the x and y axes whereas the values of the $T$ parameter are shown by the color code.  In plotting this figure, we have varied the masses of the scalars $A_1,A_2,S_1,S_2$ and $\eta^\pm$ in the range $[1,2]$ TeV. 
 In the right panel of Fig.~\ref{figSTU}, we have shown the correlation of the mass of the charged scalar, $m_{\eta^\pm}$ against the mass of the lightest CP even scalar, $m_{S_1}$, for the points that satisfy experimental constraints on the three oblique parameters. {Note that the lightest CP even dark scalar $S_1$ can have masses in the TeV range, the exact allowed range being dependent on the mass of the charged scalar $\eta^\pm$. }
 
 %From this figure, one can see that the values of $m_{S_1}$ that satisfy the constraints on oblique parameters can also accommodate the observed DM relic density, as will be seen in Section-\ref{DM}.

\subsection{Higgs diphoton decay rate} \label{sec.Higgsdiphoton} %\noindent 

The presence of {an} extra charged scalar also means that the decay rate for the $h\rightarrow \gamma \gamma $ process gets modified. This in our case, takes the form:
\begin{equation}
\Gamma (h\rightarrow \gamma \gamma )=\dfrac{\alpha _{em}^{2}m_{h}^{3}}{%
256\pi ^{3}v^{2}}\left\vert \sum_{f}a_{hff}N_{C}Q_{f}^{2}F_{1/2}(\rho
_{f})+a_{hWW}F_{1}(\rho _{W})+\frac{C_{h\eta^{\pm }\eta^{\mp }}v}{2m_{\eta^{\pm }}^{2}%
}F_{0}(\rho _{\eta^{\pm }})\right\vert ^{2},
\end{equation}%
where {$\rho_{f}=\frac{m_{h}^{2}}{4M_{f}^{2}}$, $\rho_{W}=\frac{m_{h}^{2}}{4M_{W}^{2}}$ and $\rho_{\eta^\pm}=\frac{m_{h}^{2}}{4M_{\eta^\pm}^{2}}$,} $\alpha _{em}$ is the fine structure constant; $%
N_{C}$ is the color factor ($N_{C}=1$ for leptons and $N_{C}=3$ for quarks)
and $Q_{f}$ is the electric charge of the fermion in the loop. From the
fermion-loop contributions, we only consider the dominant top quark term.
$C_{h\eta^{\pm }\eta^{\mp }}$ is the trilinear coupling
between the SM-like Higgs and a pair of charged Higges, whereas $a_{htt}$
and $a_{hWW}$ are the deviation factors from the SM Higgs-top quark coupling
and the SM Higgs-W gauge boson coupling, respectively (in the SM these
factors are unity). Such deviation factors are close to unity in our model,
which is a consequence of the fact that the singlet scalar field $\sigma $
acquires a VEV much larger than the electroweak symmetry-breaking scale.

Moreover, $F_{1/2}(z)$ and $F_{1}(z)$ are the dimensionless loop factors
for spin-$1/2$ and spin-$1$ particles running in the internal lines of the
loops. They are given by: 
\begin{align}
F_{1/2}(z)& =2(z+(z-1)f(z))z^{-2}, \\
F_{1}(z)& =-(2z^{2}+3z+3(2z-1)f(z))z^{-2}, \\
F_{0}(z)& =-(z-f(z))z^{-2},
\end{align}%
with 
\begin{equation}
f(z)=\left\{ 
\begin{array}{lcc}
\arcsin ^{2}\sqrt{z} & \text{for} & z\leq 1, \\ 
&  &  \\ 
-\frac{1}{4}\left( \ln \left( \frac{1+\sqrt{1-z^{-1}}}{1-\sqrt{1-z^{-1}}%
-i\pi }\right) ^{2}\right) & \text{for} & z>1 . \\ 
&  & 
\end{array}%
\right.
\end{equation}
In order to study the implications of our model in the decay of the $126$
GeV Higgs into a photon pair, one introduces the Higgs diphoton signal
strength $R_{\gamma \gamma }$, which is defined as
\begin{equation}
R_{\gamma \gamma }=\frac{\sigma (pp\rightarrow h)\Gamma (h\rightarrow \gamma
\gamma )}{\sigma (pp\rightarrow h)_{SM}\Gamma (h\rightarrow \gamma \gamma
)_{SM}}\simeq a_{htt}^{2}\frac{\Gamma (h\rightarrow \gamma \gamma )}{\Gamma
(h\rightarrow \gamma \gamma )_{SM}}.  \label{eqn:hgg}
\end{equation}%
That Higgs diphoton signal strength normalizes the $\gamma \gamma $ signal
predicted by our model in relation to the one given by the SM. Here we have
used the fact that in our model, single Higgs production is  dominated
by gluon fusion {into a top quark loop} as in the SM.

The ratio $R_{\gamma \gamma }$ has been measured by CMS and ATLAS
collaborations with the best fit signals \cite{10.1007/978-981-19-2354-8_33,ATLAS:2022tnm}
: 
\begin{equation}
R_{\gamma \gamma }^{CMS}=1.02_{-0.09}^{+0.11}\quad \text{and}\quad R_{\gamma
\gamma }^{ATLAS}=1.04_{-0.09}^{+0.10}.  \label{eqn:rgg}
\end{equation}
Figure \ref{Diphoton} displays the Higgs diphoton signal strength as a function of the electrically charged scalar mass, for different values of the trilinear scalar coupling $C_{h\eta^{\pm }\eta^{\mp }}$ set to be equal to $-1$ TeV, $-750$ GeV and $-500$ GeV, $1$ TeV, $750$ GeV and $500$ GeV, in the gray, purple, red, black, blue and magenta curves, respectively. The horizontal lines correspond to the $1\sigma$ CMS bounds on the Higgs diphoton signal strength. %.{\red (Why different colors for upper and lower bounds?. There is no reason for that, is just a matter of presentation)}. Figure updated.
This figure clearly shows that our model can successfully accommodate the current Higgs diphoton decay rate constraints. {One can see that the lower bound on charged scalar mass is around $300$ GeV for the value trilinear coupling $C_{h\eta^{\pm }\eta^{\mp }} = 1$ TeV. This lower bound increases further to $\sim 430$ GeV for $C_{h\eta^{\pm }\eta^{\mp }} = 500$ GeV.}

\begin{figure}[tbp]
\centering
\includegraphics[width=8cm, height=6cm]{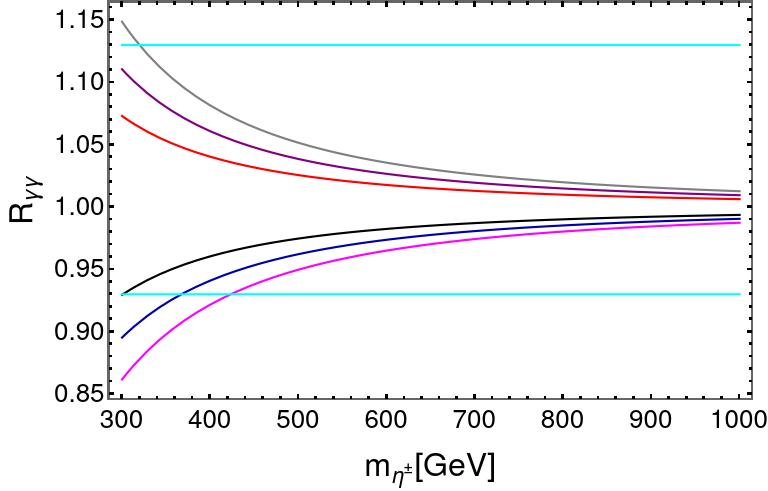}
\caption{Higgs diphoton signal strength as a function of the charged scalar
	mass for different values of the trilinear scalar coupling $C_{h\eta^{\pm }\eta^{\mp }}$. The gray, purple, red, black, blue and magenta curves correspond to values of the values of the trilinear scalar coupling $C_{h\eta^{\pm }\eta^{\mp }}$ equal to $-1$ TeV, $-750$ GeV and $-500$ GeV, $1$ TeV, $750$ GeV and $500$ GeV, respectively. %{\red(Fig x axis $m_{H^\pm} \rightarrow m_{\eta^\pm}$). Done}
 }
\label{Diphoton}
\end{figure}

\subsection{Charged scalar pair production}

\begin{figure}[tbh]
\centering
\includegraphics[width=0.5\textwidth]{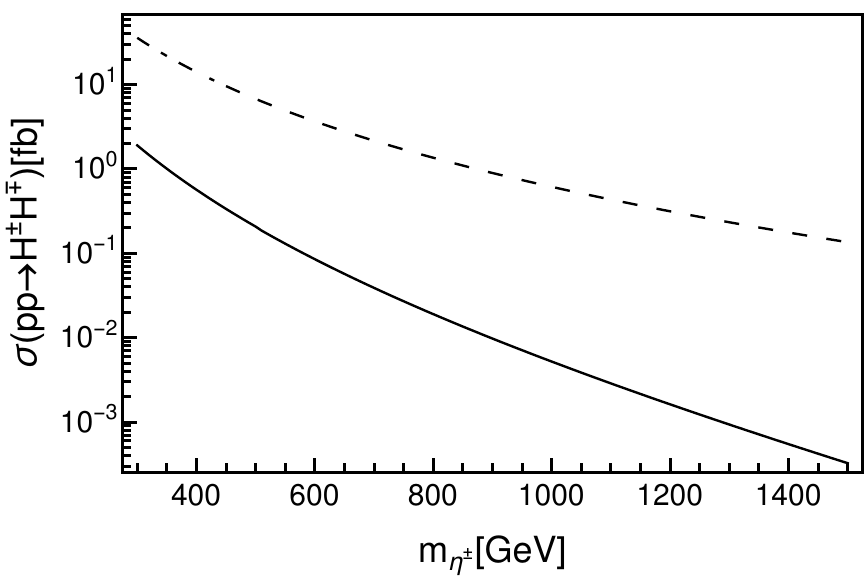}
\caption{Total cross section for the charged scalar pair production via
Drell-Yan mechanism at a proton-proton collider for $\protect\sqrt{S}=14$
TeV (continuous line) and $\protect\sqrt{S}=100$ TeV (dashed-line) as a
function of the charged scalar mass $m_{\protect\eta^\pm}$.} %{\red(fig axes $H^\pm \rightarrow \eta^\pm$)}. Done}
\label{pptoHpHn}
\end{figure}

Next we discuss the production of a scalar pair $\eta^{+}\eta^{-}$ at a
proton-proton collider. Such production mechanism at the LHC is dominated by
Drell-Yan annihilation. Fig.~\ref{pptoHpHn} displays the total cross
section for the charged scalar pair production via Drell-Yan mechanism at a
proton-proton collider for $\sqrt{S}=14$ TeV (continuous line) and $\sqrt{S}%
=100$ TeV (dashed-line) as a function of the charged scalar mass $m_{\eta^{\pm }}$. {This figure} shows that the total cross section for the scalar pair
production at a proton-proton collider ranges from $2$ fb up to $10^{-4}$
fb, for $m_{\eta^\pm} \sim 300 - 1000$ GeV and $\sqrt{S}=14$ TeV.
This total
production cross section is enhanced when the proton-proton center of mass
energy is increased and takes values ranging from $40$ fb up to about $0.1$
fb for $\sqrt{S}=100$ TeV.  For an LHC integrated luminosity of $100$ fb $^{-1}$, we get $200$ events per year. This implies $4\times 10^{-4}$
events per minute for pair production of charged scalars of mass $300$ GeV at $%
\sqrt{S}=14$ TeV. This number of
events is well below the experimental upper limit of $5$ scalars per minute
showing that charged scalars of mass $300$ GeV and above are allowed by the bounds from the LHC
searches. Moreover such values of the charged scalar mass are also allowed by the Higgs diphoton decay rate constraints as seen in the previous section. 

\VKN{Once 
$\eta^{+}$  is produced at the collider, it can decay either to 
a $W^{+}$ boson and a neutral scalar, i.e: $\eta^{+} \rightarrow W^{+} + S_i~(\textrm{or}~A_i)$ or into a charged lepton plus heavy
neutrino $\eta^{+} \rightarrow e^{+} + N$,  depending on the kinematic configuration. If $m_{W^{+}} > m_{\eta^{+}}-m_{S_i/A_i}$ the off-shell $W^{+}$ will %additionally 
decay into two jets or to a lepton plus neutrino, i.e: $\eta^{+}\rightarrow 2j + S_i~(\textrm{or}~A_i)$ or $\eta^{+}\rightarrow l+v + S_i~(\textrm{or}~A_i)$. Note that the DM candidate or the light neutrino in the final state will appear as a missing energy.
These possible final states: $W^{+}+S$, $2j+S$, $l+\nu+S$, $e^{+}+N$ %could represent 
correspond to interesting signatures that can be tested at colliders, but a detailed analysis in this direction falls beyond the scope of this work. }

 \subsection{Leptogenesis}\label{leptogenesis}

As mentioned before, the SM {does not provide} an explanation for the observed
baryon asymmetry of the Universe. The asymmetry can be expressed in terms of a parameter $\eta$ 
defined as $\eta = \frac{\eta_B - \eta_{\bar{B}}}{\eta_\gamma}$ , where $\eta_B$ , $\eta_{\bar{B}}$ and $\eta_\gamma$ are the number densities of baryons, antibaryons
and photons, respectively. The combined analysis of the data from measurements of the cosmic microwave
background and large-scale structure indicates $\eta \sim 6.1 \times 10^{-10}$~\cite{Planck:2018vyg,ParticleDataGroup:2022pth}. One appealing solution to this observed matter-antimatter asymmetry is the mechanism of leptogenesis where the out-of-equilibrium lepton number violating decay of the heavy states gives rise to a lepton asymmetry, which in turn is converted into a baryon asymmetry via the non-perturbative sphaleron processes~\cite{Fukugita:1986hr,Kolb:1990vq}.
In the traditional vanilla leptogenesis with type-I seesaw, the requirement of satisfying the correct active neutrino masses leads to a lower bound on the lightest heavy neutrino mass $\sim 10^{8} - 10^{10}$ GeV~\cite{Davidson:2002qv}. On the other hand, in our model, it is possible to lower the scale of leptogenesis much further since the light neutrino masses are also loop-suppressed. The authors of reference \cite{Hugle:2018qbw}  have studied low-scale leptogenesis in the context of the scotogenic model, where it was shown that in the version of scotogenic model with three heavy Majorana neutrinos, the scale of the decaying neutrino could be as low as 10 TeV (This is not possible if there are only two heavy neutrinos, in which case the decaying Majorana fermion should be as heavy as $10^8 - 10^{10}$ GeV.). We extend the analysis to our model and as we will see, successful leptogenesis can be realized for heavy neutrino masses as low as $10$ TeV.

Assuming that the CP asymmetry is generated by the decay of the lightest heavy neutrino $N_1$ {to the SM leptons and the components of the dark scalar $\eta$,} the relevant Boltzmann equations are given as:
\begin{equation}
    \frac{dN_{N_1}}{dz_1} = -D_1(N_{N_1} - N_{N1}^{eq}),\label{boltz1}
\end{equation}
\begin{equation}
    \frac{d N_{B-L}}{dz_1} = -\epsilon_1 D_1(N_{N_1} - N_{N1}^{eq}) - W^{tot} N_{B-L},\label{boltz2}
\end{equation}
where {$z_1 = m_{\widetilde{N}_1}/T$}. The CP asymmetry parameter $\epsilon_1$ is given by
\begin{equation}
     \epsilon_1 = \frac{1}{8 \pi ( {y^{(N)}}^\dag y^{(N)} )_{11} } \sum_{j = 2,3} \textrm{Im}~ [( {y^{(N)}}^\dag y^{(N)} )^2_{j1}] \frac{1}{\sqrt{r_{j1}}} F(r_{j1}, \eta_1),
\end{equation}
where the function $F(r_{j1}, \eta_1)$ is defined as~\cite{Buchmuller:2004nz,Davidson:2008bu,Hugle:2018qbw}
\begin{equation}
    F(r_{j1}, \eta_1) = \sqrt{r_{j1}} \Big[ f(r_{j1},\eta_1) - \frac{\sqrt{r_{j1}}}{r_{j1}-1}  (1-\eta_1)^2\Big],
\end{equation}
with $f(r_{j1}, \eta_1)$ given as
\begin{equation}
f(r_{j1},\eta_1) = \sqrt{r_{j1}} \Big[ 1 + \frac{1 - 2\eta_1 + r_{j1}}{(1-\eta_1)^2} \textrm{ln}\Big(\frac{r_{j1} - \eta_1^2}{1 - 2\eta_1 + r_{j1}} \Big) \Big].
\end{equation}
The parameters $r_{j1}$ and $\eta_1$ are defined as
{$$r_{j1} = \frac{m_{\widetilde{N}_j}^2}{m_{\widetilde{N}_1}^2}, \,\, \eta_1 = \frac{m_\eta^2}{m_{\widetilde{N}_1}^2}.$$}
{The  coefficient $D_1$  in Eqns.\ref{boltz1} and \ref{boltz2} depend on $z_1$ and is given by : }
\begin{equation}
  D_1 =  K_1 z_1 \frac{\mathcal{K}_1(z_1)}{\mathcal{K}(z_1)}, \,\,\, \textrm{where}
\end{equation}
 \begin{equation}
    K_1 = \frac{\Gamma_1}{ H(z_1 = 1)} \,, \,\,\,\, \Gamma_1 = \frac{M_1}{8\pi} ( {y^{(N)}}^\dag y^{(N)} )_{11} (1-\eta_1)^2, \,\,\,\, H = \sqrt{\frac{4 \pi^3 g^*}{45}}\frac{T^2}{M_{Pl}} = \frac{H(z_1=1)}{z_1^2}.
\end{equation}
\begin{equation}
    W^{tot} = W_1 + \Delta W, \,\,\, \textrm{where}
\end{equation}
\begin{equation}
    W_1 = \frac{1}{4} K_1 z_1^3 \mathcal{K}_1(z_1) \, , \,\,\, \Delta W = \frac{\Gamma_{\Delta L = 2}}{H z_1} = \frac{36 \sqrt{5} M_{Pl} }{\pi^{1/2} g_l \sqrt{g^*} M_1 z_1^2} \textrm{Tr}[{y^{(N)}}^T y^{(N)} ({y^{(N)}}^T y^{(N)})^\dag]\,\,\,\, \textrm{and,}
\end{equation}
\begin{equation}
    N_{N_1}^{eq} = \frac{z_1^2}{2} \kappa_2(z_1).
\end{equation}
In the above equations, {$\mathcal{K}_i(z_1)$ with $i=1,2$ are the modified Bessel functions of the second kind}, $g_l = 2$ counts the effective degrees of freedom per active neutrino, $g^* = 141$ denotes the total number of degrees of freedom and $M_{Pl}$ is the Planck scale.

\begin{figure}[tbp]
\centering
\includegraphics[width=0.45\textwidth]{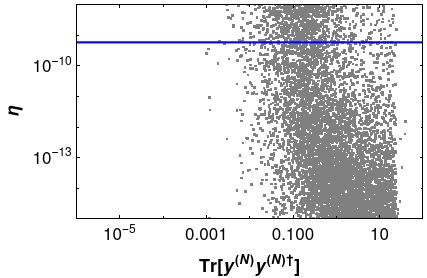}
\includegraphics[width=0.45\textwidth]{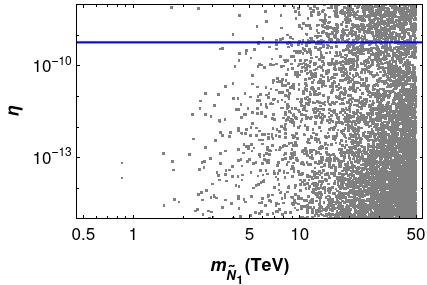}
\caption{The baryon asymmetry parameter $\eta$ as a function of Tr$[{y^{(N)}}^\dag y^{(N)} ]$ (left) and the lightest heavy neutrino mass $M_1$ (right), respectively. We have used the modified CI parametrization for $y^{(N)}$ as given in Eq.~\ref{CI} to generate this figure. Refer to the text for details.}
\label{Lepto1}
\end{figure}

We solve Eqs.\ref{boltz1} and \ref{boltz2} numerically to calculate the final $B-L$ asymmetry $N_{B-L}^f$ which in turn can be converted into baryon asymmetry $\eta$ as $\eta = C N_{B-L}^f$, where the conversion factor $C \approx 0.006$ \cite{Hugle:2018qbw}. The results of our numerical analysis are shown in Fig.~\ref{Lepto1}, where we have plotted the calculated baryon asymmetry $\eta$ as a function of Tr$[{y^{(N)}}^\dag y^{(N)} ]$ (left) and the lightest heavy neutrino mass $M_1$ (right), respectively. As in section-\ref{seclfv}, we have used the modified CI parametrization for $y^{(N)}$ as given in Eq.~\ref{CI} to generate this figure.  The masses of the scalars $\eta^\pm$, $S_1$ and $S_2$ are fixed to be 700 GeV, 500 GeV and 750 GeV respectively, whereas the masses of the two CP odd scalars $A_1$ and $A_2$ are varied in the ranges $[m_{S_1},m_{S_1}+50~\textrm{GeV}]$ and $[m_{S_2},m_{S_2}+50~\textrm{GeV}]$ respectively, the scalar mixing angles entering the expression for $\Lambda_k$ are varied in the range $[0,1]$, the two heavier Majorana neutrino masses $ m_{\tilde{N}_2}$ and $m_{\tilde{N}_3}$ are varied in ranges $[m_{\tilde{N}_1}, 2000~\textrm{TeV}]$, the active light neutrino mixing angles, mass-squared differences and CP phase are varied in the $3\sigma$ ranges~\cite{deSalas:2020pgw} and the active neutrino Majorana phases are varied in the range $[0,\pi]$. For the orthogonal matrix $R$ in Eqn.~\ref{CI}, the complex mixing angles $z_1$ and $z_2$ are taken to be $0$ whereas $z_3$ is taken as $x_3 - I x_3$ with $x_3$ varying in the range $[0,50]$. The blue horizontal line in Fig.~\ref{Lepto1} corresponds to the observed asymmetry. \VKN{All the displayed points are allowed by the bounds from perturbativity on $y^{(N)}$.} From these two figures, one can clearly see that our model can successfully accommodate the observed baryon asymmetry of the universe for $M_1 \gsim 10$ TeV and Tr$[{y^{(N)}}^\dag y^{(N)} ] \gsim 0.01$.

\subsection{Dark matter}\label{DM}
%In this section, we delve into a detailed examination of the phenomenology of the DM candidates. Our primary interest lies in understanding the consequences of the genesis of the relic density via thermal decoupling in the early Universe, commonly referred to as the freeze-out mechanism. 

In this section, we discuss the DM phenomenology. In particular, we focus on the consequences of the genesis of the relic density via thermal decoupling in the early Universe, commonly referred to as the freeze-out mechanism. The lightest of the fields that are odd under the residual $Z_2$ symmetry - namely, the scalar mass eigenstates $S_1, S_2, A_1$ and $A_2$, or the lightest right-handed Majorana neutrino - can be the stable DM candidate in our model.

%The residual $\mathrm{Z}_2$  symmetry which emerges following the spontaneous breaking of the UV $\mathrm{Z}_4$ symmetry stabilizes the lightest of the neutral fields with charge $\pm 1$ under $\mathrm{Z}_4$ transformations, namely, one of the scalar mass eigenstates $S_1, S_2, A_1$ and $A_2$, or the lightest right-handed Majorana neutrino. 

Note that even though one of the three Majorana right-handed neutrinos $N_i$ could be a DM candidate, such a scenario will be incompatible with low-scale leptogenesis. This is because as mentioned earlier, to have successful low-scale leptogenesis, the neutrino sector should consist of three heavy Majorana neutrinos, the lightest of which decays to produce the required lepton asymmetry\footnote{Leaving leptogenesis aside, an estimation based on Ref. \cite{Bernal:2017xat} gives cosmologically acceptable results for DM relic density for a Yukawa coupling equal to unity, a $400$ GeV mass for the fermionic DM candidate and $m_{\eta }\sim 1.6$ TeV.}. Thus we focus only on the scalar DM candidate and its phenomenology. Our approach involves simplifying the analysis by considering the masses of the exotic fermions to be much greater than that of the scalar particles, thus effectively decoupling their interaction with the DM candidate. %during and after the decoupling process occurs.

Consequently, we consider only the scalar potential presented in the App. \ref{Appendix} {for DM phenomenology}. We present a few illustrative numerical results that underscore the
viability and scope of 
the DM-related predictions derived from the model. These results stem from a systematic exploration of a subset of the model's parameter space, in which we minimize a logarithmic likelihood function for the DM abundance\footnote{In the high statistic limit this procedure proves as useful as minimizing a $\chi^2$-function defined accordingly.} as measured by Planck \cite{Planck:2018vyg}, {which states $\Omega_c h^2 =  0.11933 \pm 0.00091$ at $68 \%$ confidence level, where $\Omega_c$ is the ratio of cold DM energy density and the critical density of the Universe, and $h=67.27 \pm 0.60$ quantifies the Hubble constant as defined by $H_0=h $ km s$^{-1}$ Mpc$^{-1}$}.

Conducting a thorough numerical scan within the recently described regime poses a formidable 16-dimensional optimization problem. We have chosen to take 13 coupling constants (comprising 11 quartic ones and the trilinear coupling constants $C_1$ and $C_2$) alongside 3 free $\mu$ parameters as independent variables. Consequently, masses, vacuum expectation values, as well as mixing angles between interaction eigenstates, were considered dependent variables, as described in the App. \ref{Appendix}.

To simplify our analysis, it is important to note that the self-interactions governed by $\lambda_2$, $\lambda_3$, and $\lambda_4$ do not directly impact the Boltzmann equations. Instead, their influence is mediated through the masses of particles in the dark sector. Consequently, we set these parameters to $0.1$. To suppress mixing between $H_1$ and $H_2$ and ensure that $H_1$ retains properties akin to the Higgs in the SM, we set the constant $\lambda_7$ to 0.001. This choice does not diminish the role of $H_2$ as a scalar portal in the decoupling process. Similarly,  we fixed $\lambda_{10}$ at an arbitrary value of $0.001$ since it does not appear in the mass matrix.  Its negligible impact on the solutions to Boltzmann equations was verified afterward.

For simplicity, we choose $C_1=0$. Then, we selected the $\mu$ parameters and the coupling constant $C_2$ in two distinctive scenarios: one where the dark matter (DM) is CP-even and another where it is CP-odd. For the remaining six coupling constants, we chose values within the range of 0.001 to 10, ensuring that this parameter selection adheres to the theoretical requirements of vacuum stability and perturbativity.

We use the differential evolution algorithm \cite{Storn97} from the \texttt{scipy} package \cite{virtanen2020scipy} for \texttt{Python} to search for points that saturate the totality of the DM budget, solving the Boltzmann equation using \texttt{micrOMEGAs 5.3.41} \cite{BELANGER2018173}. The entire model implementation was carried out using \texttt{FeynRules} \cite{Alloul_2014}. 

The annihilation channels contributing to the Boltzmann equations are mainly the $s$-channel annihilation into the SM gauge bosons and fermions through the two Higgs portals. The seagull, contact four-point interaction coming from the covariant derivative of the doublet scalar $\eta$ is a relevant extra annihilation channel depending on the magnitude of the mixing between the doublet and singlet states. The coannihilations between the non-stable scalars and the DM candidate are also relevant in our particular scan because the mass differences lie below $\sim 15\%$.

This scenario can be proved with deep underground experiments like XENONnT \cite{XENON:2018voc,XENON:2023cxc} or LZ \cite{LZ:2022lsv}, capable of measuring the recoil energy of elastic
scatterings of DM  off nucleons, which in our model are mediated by the Higgs particle and its partner. The $Z$ boson does not act as a mediator because the trilinear interactions of dark scalars with the $Z$ are
\begin{eqnarray}
\mathcal{L}&\supset&\frac{M_{Z}}{v}\left( \cos \theta _{S}S_{1}-\sin \theta _{S}S_{2}\right)
\left( \cos \theta _{A}\partial _{\mu }A_{1}-\sin \theta _{A}\partial _{\mu
}A_{2}\right) Z^{\mu }\notag\\
&&-\frac{M_{Z}}{v}\left( \cos \theta _{A}A_{1}-\sin \theta _{A}A_{2}\right)
\partial _{\mu }\left( \cos \theta _{S}\partial _{\mu }S_{1}-\sin \theta
_{S}\partial _{\mu }S_{2}\right) Z^{\mu },
%&&i\frac{M_{Z}}{v}\left( \cos ^{2}\theta _{W}-\sin ^{2}\theta _{W}\right)\left( H^{+}\partial _{\mu }H^{-}-H^{-}\partial _{\mu }H^{+}\right) Z^{\mu }\notag\\
\end{eqnarray}
which are irrelevant for elastic DM-nucleon scattering. 
\begin{figure}[t]
\centering
\includegraphics[width=0.48\textwidth]{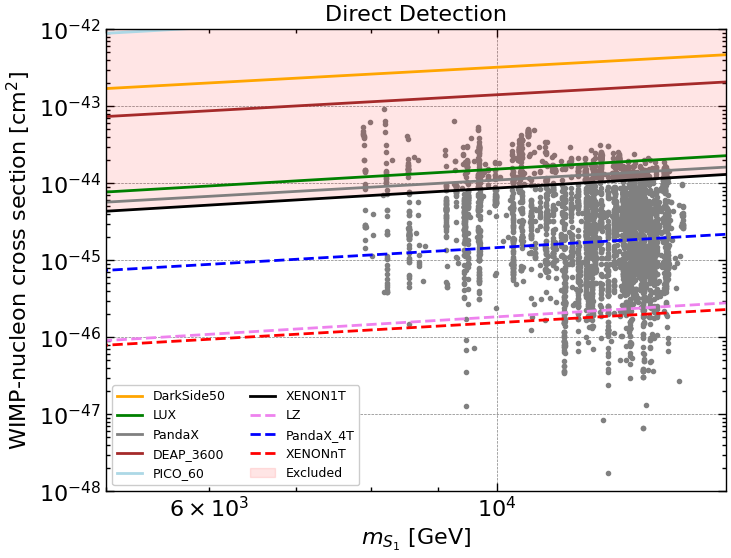}
\includegraphics[width=0.48\textwidth]{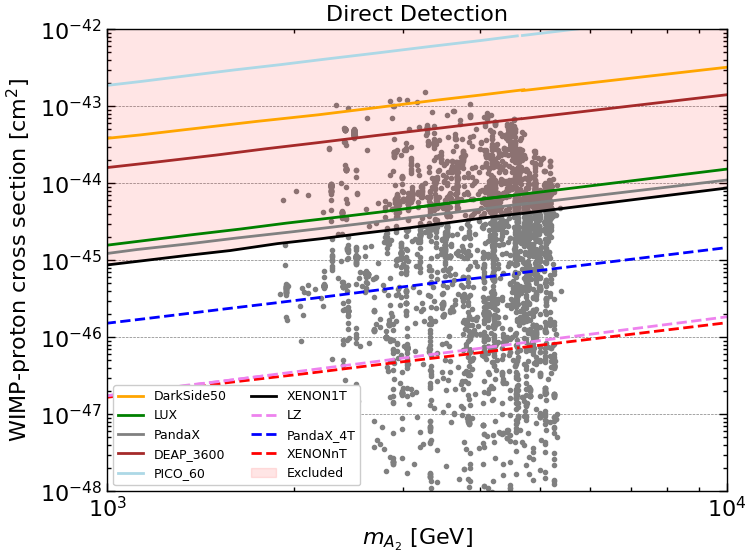}
\caption{DM-nucleon spin-independent elastic scattering cross-section as a function of the DM mass. The continuous and dashed lines represent upper limits bounds and prospects for different direct detection experiments, adapted from \cite{Akerib:2022ort} (left) CP-even candidate, $\mu_\sigma = 2000$ GeV, $\mu_\eta = 1500 $ GeV, $\mu_\xi =1000$ GeV, $C_2=-100$ GeV. (right) CP-odd candidate, $\mu_\sigma = 1000$ GeV, $\mu_\eta = 1500 $ GeV, $\mu_\xi =2000$ GeV, $C_2=100$ GeV}
\label{dd}
\end{figure}
The constraints obtained are shown in Fig. \ref{dd}, the left and right plots corresponding to CP-even and CP-odd DM candidates respectively. All points displayed saturate the cosmological DM abundance with $\sim 10\%$ tolerance, i.e., $\Omega_{DM} h^2=0.12\pm 0.01$, and have dark matter masses of around $10$ TeV. The vertical axis represents the DM-nucleon spin-independent elastic scattering cross-section, and the horizontal axis represents the DM mass. Current measurements of the WIMP-nucleon cross-section exclude points overlapping with the faint red region. The model can generically accommodate direct detection signals across 5 orders of magnitude, so we have points already excluded by XENON1T, others observable in next-generation experiments, and others down to one order of magnitude below the expectations for those detectors, falling below prospects of detection even for the DARWIN \cite{DARWIN:2016hyl} collaboration. 
\begin{figure}[t]
\centering
\includegraphics[width=0.48\textwidth]{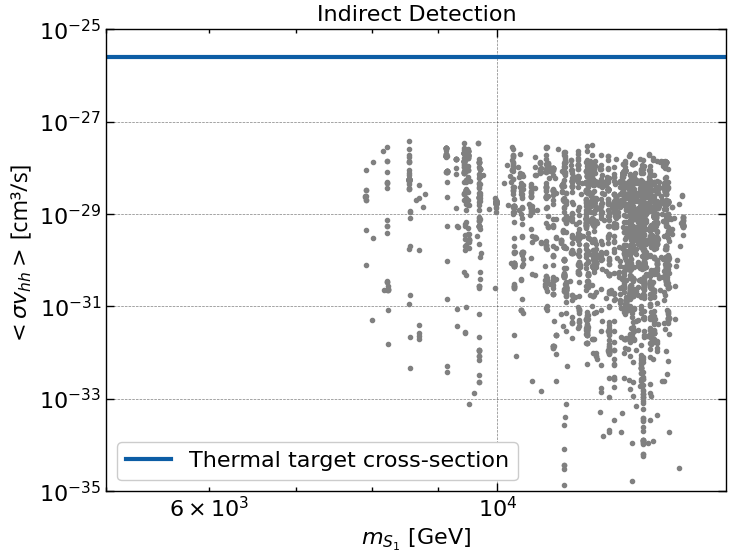}
\includegraphics[width=0.48\textwidth]{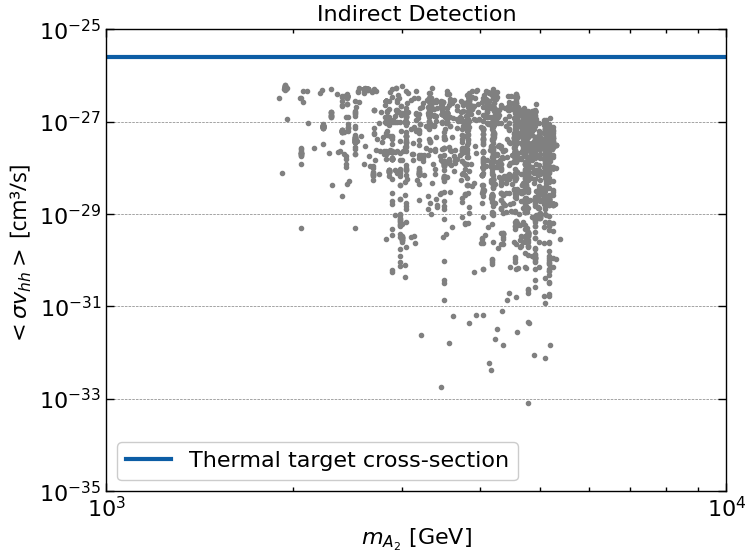}
\caption{Thermally averaged DM annihilation cross-section into a Higgs pair, in the non-relativistic approximation, i.e., keeping just the $s$-wave. (left) $\mu_\sigma = 2000$ GeV, $\mu_\eta = 1500 $ GeV, $\mu_\xi =1000$ GeV, $C_2=-100$ GeV. (right)  $\mu_\sigma = 1000$ GeV, $\mu_\eta = 1500 $ GeV, $\mu_\xi =2000$ GeV, $C_2=100$ GeV}
\label{id}
\end{figure}

We now consider the possibility of discovering or ruling out the DM candidate through the identification of the products of its annihilation in the galactic halo. To do so we project the points allowed by the considerations exposed for Fig. \ref{dd} in a plane relevant for indirect detection, in Fig. \ref{id}, where we represent DM annihilation cross-section into a pair of Higgs bosons in the vertical axis, as a function of DM mass in the horizontal axis. The blue line represents the canonical value $\langle \sigma v \rangle \approx 2\times10^{-26}$ cm$^3$ s$^{-1}$, which lies in the ballpark of sensitivity of most detectors of cosmic rays as Fermi-LAT \cite{Fermi-LAT:2016uux}, AMS \cite{Kopp:2013eka}, or CTA \cite{Duangchan:2022jqn}\footnote{The upper bounds constructed from the data of these collaborations are model dependent. For instance, the presence of a Higgs partner should modify the gamma-ray spectra of photons coming from DM annihilation into $h_1$ pairs, and the $h_1,h_2$ and $h_2,h_2$ final states interfere with it. A detailed analysis of cosmic ray spectra is beyond the scope of this work}. The subsequent decays of the Higgs boson into $b$ quark pairs are the ones with more constraining power for Higgs portal DM \cite{DiazSaez:2021pmg}.
Regardless of this, our model has difficulties to be identified or constrained in the present and next generation of cosmic-ray detectors. Even in the most optimistic case, the annihilation cross-section lies at least one order of magnitude below the
sensibility of these {instruments}. 

\section{Conclusions}
\label{conclusions}
We have constructed an extension of the Inert Doublet Model (IDM) that successfully accommodates the SM-charged fermion mass hierarchy, is compatible with the neutrino oscillation experimental data and can explain the observed DM relic abundance and baryon asymmetry of the universe. In addition to the SM particle content, the model contains one inert doublet and two electrically neutral singlet scalars, three right-handed Majorana neutrinos and additional charged vector-like fermions. A discrete $Z_4$ symmetry prevents tree-level masses of the active light neutrinos and the first two generations of the SM charged fermions. Consequently, their masses are generated at one-loop level, thereby providing an explanation to the huge hierarchy in the fermion mass spectrum. {Of the three extra scalars, the doublet and one singlet (which we collectively refer to as the dark scalars) are odd under the remnant preserved $Z_2$ symmetry and hence do not acquire $vev$s providing a stable DM candidate whereas the other scalar is even under $Z_2$ and acquires a $vev$ that preserves the $Z_2$ symmetry. Thus, an interesting feature of the model is that in addition to explaining the observed DM relic density, the dark scalars also (i) contribute to the one-loop generation of the fermion masses thus explaining light neutrino masses and fermion mass hierarchy, (ii) contribute to the charged lepton flavor violation, (iii) participate in leptogenesis and (iv) can give rise to interesting collider signatures  - thereby providing a connection between the various phenomenological aspects of the model.}

In a large part of the parameter space, the model successfully complies with the constraints arising from the muon and the electron $g-2$  anomalies, electroweak precision observables, and Higgs diphoton decay rate. The smallness of the observed neutrino masses, %$m_{\nu }$
together with the requirement of explaining the observed baryon asymmetry of the Universe puts a lower bound of around $10$ TeV on the lightest  heavy Majorana neutrino mass. This low-scale leptogenesis scenario is possible as a consequence of the loop suppression of the active light neutrino masses, and differs from the traditional vanilla leptogenesis where the heavy neutrino mass $m_N$ is typically of the order $ 10^{8}- 10^{10}$ GeV. For this scenario to work, the Yukawa couplings have to satisfy the bound - Tr$[{y^{(N)}}^\dag y^{(N)}]\gsim 0.01$. The parameter space that we consider are within the current bounds from charged lepton flavor violation.

The constraints from the experimental values of the muon and the electron anomalous magnetic moments set the dark scalar and charged exotic lepton masses to be around the TeV scale. Moreover, our analysis of the model's implications on the Higgs diphoton decay sets a lower bound on the electrically charged scalar mass of around 300 GeV, whose exact value depends on the trilinear scalar coupling between the SM-like Higgs and the pair of charged Higgs.

As mentioned, the scalar DM candidate is stabilized by the remnant preserved $Z_{2}$ symmetry arising from the spontaneous breaking of the $Z_{4}$ symmetry. The correct relic abundance as indicated by the Planck bound is satisfied for a wide range of scalar DM masses. Part of the parameter space allowed by the constraints from the current and the future direct and indirect DM detection experiments, also fulfills all the phenomenological constraints arising from leptogenesis, neutrino oscillation data and $g-2$  experiments.

\section*{Acknowledgments}

P.E.C. thanks support by ANID-Chile Grant 21210952, and grant PIIC 2022-I, DPP UTFSM. 
 V.K.N. is supported by ANID-Chile Fondecyt Postdoctoral grant 3220005.
The authors also thank the support from ANID-Chile FONDECYT grants No. 1210131, No. 1210378,  No. 1230110 and No. 1231248, from ANID PIA/APOYO AFB230003, and from ANID Millenium Science Initiative ANID-ICN2019-044. Authors thank Luis Lavoura for pointing out a couple of misprints in the previous version of our manuscript.

\appendix

\section{Scalar potential}\label{Appendix}
The scalar potential invariant under the symmetries of the model is given
as,
\begin{align}
V=& -\mu _{\phi }^{2}(\phi ^{\dagger }\phi )+\mu _{\eta }^{2}(\eta ^{\dagger
}\eta )+\mu _{\xi }^{2}(\xi ^{\ast }\xi )-\mu _{\sigma }^{2}\sigma
^{2}+\lambda _{1}(\phi ^{\dagger }\phi )^{2}+\lambda _{2}(\eta ^{\dagger
}\eta )^{2}  \notag \\
& +\lambda _{3}(\xi ^{\ast }\xi )^{2}+\lambda _{4}\sigma ^{4}+\lambda
_{5}(\phi ^{\dagger }\phi )(\eta ^{\dagger }\eta )+\lambda _{6}(\phi
^{\dagger }\eta )(\eta ^{\dagger }\phi )+\lambda _{7}(\phi ^{\dagger }\phi
)\sigma ^{2}  \notag \\
& +\lambda _{8}(\phi ^{\dagger }\phi )(\xi ^{\ast }\xi )+\lambda _{9}(\eta
^{\dagger }\eta )\sigma ^{2}+\lambda _{10}(\eta ^{\dagger }\eta )(\xi ^{\ast
}\xi )+\lambda _{11}(\xi ^{\ast }\xi )\sigma ^{2}  \notag \\
& +\lambda _{12}\left( (\eta \phi ^{\dagger })\xi ^{\ast }\sigma +h.c\right)
+ C_{1} \left( \phi ^{\dagger }\eta  \xi +h.c \right)
+C_{2}\left( \xi ^{2}\sigma +h.c\right). \notag
\end{align}%
%{\bf (From the way it is written, it appears as if $C_1$ is complex and $C_2$ is real. Should we keep both as complex or both as real?)}
%where the $SU\left( 2\right) _{L}$ scalar doublets can be expanded as, 
%\begin{equation} \phi =\left(  \begin{array}{c} \phi ^{+} \\  \frac{1}{\sqrt{2}}\left( v+\phi _{R}^{0}+i\phi _{I}^{0}\right)  \end{array}% \right) ,\qquad \eta =\left(  \begin{array}{c} \eta ^{+} \\  \frac{1}{\sqrt{2}}\left( \eta _{R}^{0}+i\eta _{I}^{0}\right)  \end{array}% \right) . \end{equation}%

The minimization conditions resulting from the scalar potential given above are,
\begin{eqnarray}
\mu _{\phi }^{2} &=&\lambda _{1}v^{2}+\frac{1}{2}\lambda _{7}v_{\sigma }^{2},
\notag \\
\mu _{\sigma }^{2} &=&\lambda _{4}v_{\sigma }^{2}+\frac{1}{2}\lambda
_{7}v^{2}.\ 
\end{eqnarray}%
From the analysis of the scalar potential, it follows that $\phi _{I}^{0}$
is the massless CP odd scalar field corresponding to the SM neutral
Goldstone boson. The remaining CP odd neutral scalar fields, i.e., $\eta _{I}$ and $\xi
_{I}$ mix among themselves and the corresponding  mass-squared matrix in the basis $%
\left( \eta _{I},\xi _{I}\right) $ has the form, 
\begin{equation}
M_{A}^{2}=\left( 
\begin{array}{cc}
\frac{1}{2}\left( 2\mu _{\eta }^{2}+\lambda _{5}v^{2}+\lambda _{9}v_{\sigma
}^{2}\right)  & \frac{1}{2}\lambda _{12}vv_{\sigma }-\frac{1}{\sqrt{2}}C_{1}v
\\ 
\frac{1}{2}\lambda _{12}vv_{\sigma }-\frac{1}{\sqrt{2}}C_{1}v & \frac{1}{2}%
\left( -2\sqrt{2}C_2v_{\sigma }+2\mu _{\xi }^{2}+\lambda _{8}v^{2}+\lambda
_{11}v_{\sigma }^{2}\right)  \\ 
& 
\end{array}%
\right) .
\end{equation}
This matrix can be diagonalized as,
\begin{eqnarray}
R_{A}^{T}M_{A}^{2}R_{A} &=&\left( 
\begin{array}{cc}
\frac{TrM_{A}^{2}}{2}+\frac{1}{2}\sqrt{\left( TrM_{A}^{2}\right) ^{2}-4\det
M_{A}^{2}} & 0 \\ 
0 & \frac{TrM_{A}^{2}}{2}-\frac{1}{2}\sqrt{\left( TrM_{A}^{2}\right)
^{2}-4\det M_{A}^{2}}%
\end{array}%
\right) ,  \notag  \label{eq:Theta-S} \\
R_{A} &=&\left( 
\begin{array}{cc}
\cos \theta _{A} & -\sin \theta _{A} \\ 
\sin \theta _{A} & \cos \theta _{A}%
\end{array}%
\right) ,\hspace{0.5cm}\hspace{0.7cm}\tan 2\theta _{A}=\frac{2\left(
M_{A}^{2}\right) _{12}}{\left( M_{A}^{2}\right) _{11}-\left(
M_{A}^{2}\right) _{22}}.  \notag
\end{eqnarray}
Consequently, the physical inert CP odd neutral scalar mass eigenstates $%
A_{1,2}$ are given by:
\begin{equation}
\left( 
\begin{array}{c}
A_{1} \\ 
A_{2}%
\end{array}%
\right) =\left( 
\begin{array}{cc}
\cos \theta _{A} & \sin \theta _{A} \\ 
-\sin \theta _{A} & \cos \theta _{A}%
\end{array}%
\right) \left( 
\begin{array}{c}
\eta _{I} \\ 
\xi _{I}%
\end{array}%
\right) .\notag
\end{equation}
Due to the remnant preserved $Z_{2}$ symmetry, the CP even parts of the neutral component of $\phi$ (i.e, $\phi _{R}^{0}$) and $\tilde{\sigma}$ do not mix
with the remaining CP even neutral scalar fields. The corresponding mass-squared matrix in the basis $\left( \phi _{R}^{0},\tilde{\sigma}\right) $ can be written as follows: 
\begin{equation*}
M_{H}^{2}=\left( 
\begin{array}{cc}
2\lambda _{1}v^{2} & \lambda _{7}vv_{\sigma } \\ 
\lambda _{7}vv_{\sigma } & 2\lambda _{4}v_{\sigma }^{2}  
\end{array}%
\right) ,
\end{equation*}%
where in the decoupling limit $\lambda _{7}\rightarrow 0$, $\phi _{R}^{0}$
will correspond to the  $126$ GeV SM Higgs boson. The remaining CP even
neutral scalar fields charged under the remnant $Z_{2}$ symmetry, i.e., $%
\eta _{R}$ and $\xi _{R}$ mix among themselves and the corresponding mass-squared
matrix in the basis $\left( \eta _{R},\xi _{R}\right) $ has the form,
\begin{equation}
M_{S}^{2}=\left( 
\begin{array}{cc}
\frac{1}{2}\left( 2\mu _{\eta }^{2}+\lambda _{5}v^{2}+\lambda _{9}v_{\sigma
}^{2}\right)  & \frac{1}{2}\lambda _{12}vv_{\sigma }+\frac{1}{\sqrt{2}}C_{1}v
\\ 
\frac{1}{2}\lambda _{12}vv_{\sigma }+\frac{1}{\sqrt{2}}C_{1}v & \sqrt{2}%
C_2v_{\sigma }+\mu _{\xi }^{2}+\frac{\lambda _{8}v^{2}}{2}+\frac{1}{2}\lambda
_{11}v_{\sigma }^{2} 
\end{array}%
\right) .
\end{equation}
The squared mass matrix for the inert CP even scalars can be diagonalized as,
\begin{eqnarray}
R_{S}^{T}M_{S}^{2}R_{S} &=&\left( 
\begin{array}{cc}
\frac{TrM_{S}^{2}}{2}+\frac{1}{2}\sqrt{\left( TrM_{S}^{2}\right) ^{2}-4\det
M_{S}^{2}} & 0 \\ 
0 & \frac{TrM_{S}^{2}}{2}-\frac{1}{2}\sqrt{\left( TrM_{S}^{2}\right)
^{2}-4\det M_{S}^{2}}%
\end{array}%
\right) ,  \notag \\
R_{S} &=&\left( 
\begin{array}{cc}
\cos \theta _{S} & -\sin \theta _{S} \\ 
\sin \theta _{S} & \cos \theta _{S}%
\end{array}%
\right) ,\hspace{0.5cm}\hspace{0.7cm}\tan 2\theta _{S}=\frac{2\left(
M_{S}^{2}\right) _{12}}{\left( M_{S}^{2}\right) _{11}-\left(
M_{S}^{2}\right) _{22}}.
\end{eqnarray}
Consequently, the physical inert CP even neutral scalar mass eigenstates $%
S_{1,2}$ are given by:
\begin{equation}
\left( 
\begin{array}{c}
S_{1} \\ 
S_{2}%
\end{array}%
\right) =\left( 
\begin{array}{cc}
\cos \theta _{S} & \sin \theta _{S} \\ 
-\sin \theta _{S} & \cos \theta _{S}%
\end{array}%
\right) \left( 
\begin{array}{c}
\eta _{R} \\ 
\xi _{R}%
\end{array}%
\right) ,  \notag
\end{equation}
Finally, $\phi ^{\pm }$ correspond to the SM charged Goldstone bosons, whereas the mass of the
remaining electrically charged scalar $\eta ^{\pm }$ is given as,
\begin{equation}
m_{\eta ^{\pm }}^{2}=\frac{1}{2}\left( 2\mu _{\eta }^{2}+\lambda
_{5}v^{2}+\lambda _{9}v_{\sigma }^{2}\right).
\end{equation}

\bibliographystyle{utphys}
\bibliography{BiblioIDM17thMarch2023}

%\section*{Acknowledgments}
%\appendix
%\bibliographystyle{utphys}
%\bibliography{Biblio}

\end{document}